\documentclass[preprint,aps,prx,preprintnumbers,amsmath,amssymb,superscriptaddress,floatfix,longbibliography]{revtex4-2}

\usepackage[utf8]{inputenc}
\usepackage{graphicx}
\usepackage{color}
\usepackage{soul}
\usepackage[hidelinks]{hyperref}
\hypersetup{
 setpagesize=false,
 bookmarksnumbered=true,%
 bookmarksopen=true,%
 colorlinks=true,%
 urlcolor=blue,
 linkcolor=blue,
 citecolor=blue,
}

\usepackage{xcolor}
\usepackage{siunitx}
\usepackage{physics}
\usepackage[version=4]{mhchem}
\usepackage{geometry}

\begin{document} 

\title{Twisted Bogoliubov quasiparticles in the superconducting \texorpdfstring{\ce{NbSe2}}{NbSe2} monolayer on graphene} 

\author{Masahiro~Naritsuka}
\email[Correspondence to: ]{masahiro.naritsuka@riken.jp}
\affiliation{RIKEN Center for Emergent Matter Science, Wako, Saitama 351-0198, Japan}
\author{Tadashi~Machida}
\affiliation{RIKEN Center for Emergent Matter Science, Wako, Saitama 351-0198, Japan}
\author{Shun~Asano}
\affiliation{Department of Physics, Kyoto University, Kyoto 606-8502 Japan}
\author{Youichi~Yanase}
\affiliation{Department of Physics, Kyoto University, Kyoto 606-8502 Japan}
\author{Tetsuo~Hanaguri}
\email[Correspondence to: ]{hanaguri@riken.jp}
\affiliation{RIKEN Center for Emergent Matter Science, Wako, Saitama 351-0198, Japan}

\date{\today}

\begin{abstract}
\textbf{}
The superconducting properties of layered materials can be controlled by thinning, stacking, and twisting, demanding investigation of electronic states by spectroscopic means at the nanometer scale.
Here, we reveal the spatial variations of the electronic states in heterostructures of the superconducting monolayer \ce{NbSe2}/graphene using spectroscopic-imaging scanning tunneling microscopy.
The \ce{NbSe2} monolayer grown by molecular beam epitaxy is naturally twisted with respect to the graphene substrate and exhibits interference patterns of Bogoliubov quasiparticles twisted with respect to the \ce{NbSe2} and graphene lattices.
We find that the twisted interference patterns originate from a sextet of regions in momentum space where the Fermi surfaces of \ce{NbSe2} and graphene overlap. The Fermi surface overlap is sensitive to the twist angle, providing a knob to tune superconductivity.
\end{abstract}

\baselineskip24pt

\maketitle 

\newpage
Realizing nontrivial quantum states by processing or combining known materials is an outstanding challenge in condensed matter physics.
The layered van der Waals compounds, which exhibit various magnetic, topological, and superconducting properties, have been at the forefront of the quest for quantum phenomena in this regard~\cite{Geim2013,Novoselov2016,Liu2019}.
Many of the van der Waals compounds can be thinned down to the two-dimensional (2D) monolayer, which can be electronically distinct from the bulk crystal~\cite{Novoselov2004,Mak2010,Splendiani2010,Deng2018,Seyler2018}.
The monolayer also serves as a building block for synthesizing an artificial system by stacking.
Combining the layers with different properties can lead to synergistic or emergent phenomena such as putative topological superconductivity in superconductor/topological insulator heterostructures~\cite{Fu2008,Xu2014,Xu2015,Lupke2020,Kezilebieke2022}.
Besides stacking, weak van der Waals bonding allows twisting between adjacent layers, providing an additional knob to control electronic states~\cite{Bistritzer2011,Kennes2021}.
A noticeable example is a magic-angle twisted bilayer graphene where the band-structure control through the moir\'e band formation induces strongly correlated superconducting state~\cite{Cao2018_1,Yankowitz2019,Lu2019,Stepanov2020,Oh2021}.

To control the quantum phenomena arising from thinning, stacking, and twisting, experimental investigations of the underlying electronic states are indispensable.
This is especially true for superconducting phenomena, which are often challenging to predict theoretically.
However, advanced spectroscopic tools such as angle-resolved photoemission spectroscopy and spectroscopic-imaging scanning tunneling microscopy (SI-STM) have rarely been applied for 2D superconductors due to their low superconducting transition temperature $T_{\mathrm c}$.
In particular, the momentum-space properties of the superconducting gap have remained totally unknown.

We address this problem by investigating model twisted stacks of superconducting \ce{NbSe2} monolayers on graphene using SI-STM at an ultra-low temperature.
Bulk 2\textit{H}-\ce{NbSe2} is a well-known layered superconductor that exhibits a $3 \times 3$ charge density wave (CDW) order at \SI{30}{\K} coexisting with superconductivity below \SI{7}{\K}~\cite{Wilson2001,Moncton1975,Revolinsky1965}.
Both $3 \times 3$ CDW and superconductivity survive down to the monolayer limit, although $T_{\mathrm c}$ is reduced to about \SI{1}{\K}~\cite{Ugeda2016,Xi2016,Wang2017,Xing2017,Zhao2019,Chen2020,Dreher2021,Ganguli2022}.
Scanning tunneling spectroscopies have been performed on \ce{NbSe2} monolayers on graphene~\cite{Ugeda2016,Zhao2019,Dreher2021} or graphite~\cite{Chen2020,Ganguli2022}  and have revealed the roles of disorders~\cite{Zhao2019}, and electron correlations~\cite{Ganguli2022}.
However, the effects of stacking and twisting have yet to be elucidated.
In this work, we have succeeded in imaging the spatial variations of the superconducting gap and the Bogoliubov quasiparticles.
Using the Fourier analysis, we identify the twist-angle-dependent momentum-space regions that govern the Bogoliubov quasiparticles, providing a clue to controlling superconductivity by twist.


Twisted stacking is often achieved by the top-down stamping technique.
In the present work, we adopted the alternative bottom-up approach using molecular beam epitaxy (MBE), which naturally introduces a twist during growth.
All the SI-STM experiments were performed using a homemade dilution-refrigerator-based STM system with the lowest effective electron temperature of about \SI{90}{\milli\K}\cite{Machida2018}.
Details of the experimental procedures are given in \cite{Supplement}.

Figure~\ref{fig:1}(a) shows the STM image representing the typical surface morphology of the \ce{NbSe2} atomic layers grown on the bilayer-graphene-terminated 4\textit{H}-SiC(0001) surface.
The largest terrace corresponds to the \ce{NbSe2} monolayer region, on which small bilayer and trilayer islands are seen.
The lowest terrace is the partially exposed graphene substrate.
The atomic resolution STM topography of the \ce{NbSe2} monolayer (Fig.~\ref{fig:1}(b)) exhibits a $3 \times 3$ CDW modulation superimposed on the triangular lattice of the topmost Se atoms, being similar to the STM topography of the cleaved surface of bulk 2\textit{H}-\ce{NbSe2}.
In the graphene region, a superstructure with a periodicity of $\sim$\SI{1.8}{nm} is observed on top of the atomic corrugations (Fig.~\ref{fig:1}(c)).
This is known to be associated with the $6\sqrt{3}\times6\sqrt{3}-\mathrm{R}\ang{30}$ periodic dangling bond (DB) formation between the topmost SiC layer and the buffer layer graphene~\cite{Goler2013,Riedl2010} as schematically illustrated in Fig.~\ref{fig:1}(d).

The weak van der Waals bonding between the \ce{NbSe2} monolayer allows for a twisted stacking.
We determine the details of the twist structure by analyzing the Fourier-transformed (FT) STM topography where periodic modulations overlapping in real space can be disentangled in wave vector $\mathbf{q}$ space.
Figure~\ref{fig:1}(e) shows the FT-STM topography of the \ce{NbSe2} monolayer.
The Se-lattice Bragg peaks (blue circles), the 3$\times$3 CDW peaks (blue squares), and their higher harmonics and combinations (dashed blue symbols) are identified.
Interestingly, a set of peaks (orange triangles) appears around the origin and the Se-lattice Bragg peaks, which can not be attributed to the modulations originated in \ce{NbSe2}.

We show that these peaks arise due to the $6\sqrt{3}\times6\sqrt{3}-\mathrm{R}\ang{30}$ DB order in the underlying graphene substrate.
Figure~\ref{fig:1}(f) depicts the FT-STM topography of the nearby exposed graphene region.
Although the DB-order peaks around the origin are smeared by the inhomogeneities that govern the small-$\mathbf{q}$ region, the graphene-lattice Bragg peaks (orange circles) accompany the satellite peaks shifted by the wave vectors of the DB order (dashed orange triangle).
As shown in the insets of Fig.~\ref{fig:1}~(e and f), the non-\ce{NbSe2}-related peaks in Fig.~\ref{fig:1}(e) appear at the same positions as the satellite peaks in Fig.~\ref{fig:1}(f), indicating their DB-order origin.
Since the DB order is commensurate with the graphene lattice, we can determine the twist angle from the positions of the Se-lattice Bragg peaks and the DB-order peaks.
We define the twist angle $\theta = \ang{0}$ so that the principal lattice vector of \ce{NbSe2} and the zigzag direction of graphene are parallel (Fig.~\ref{fig:1}(d)).
The \ce{NbSe2} monolayer shown in Fig.~\ref{fig:1} (b and e) has $\theta = \ang{28}$.


To investigate the impact of the twisted stacking on the electronic state and superconductivity, we performed SI-STM in which we measure the differential conductance  $g(\mathbf{r},E=eV) \equiv \dv*{I(\mathbf{r},V)}{V}$, which reflects the local density of states (LDOS) at the lateral position $\mathbf{r}$ and energy $E$, at every pixel of the STM image.
Here, $e$ is the elementary charge, $V$ is the sample bias voltage, and $I$ is the tunneling current.
We normalize $g(\mathbf{r},eV)$ as $L(\mathbf{r},eV) \equiv g(\mathbf{r},eV)/(I/V)$ to avoid the so-called set-point effect causing a spurious spatial modulation in $g(\mathbf{r},eV)$ when the integrated LDOS up to the feedback set-point bias $V_\mathrm{set}$ is spatially inhomogeneous~\cite{Kohsaka2007}.

$L(\mathbf{r},+\SI{200}{meV})$ and its Fourier transform $L(\mathbf{q},+\SI{200}{meV})$ of the \ce{NbSe2} monolayer are shown in Fig.~\ref{fig:2}~(a and b), respectively.
$L(\mathbf{q},+\SI{200}{meV})$ exhibits peaks (red diamonds) that are not directly related to the Se lattice, $3 \times 3$ CDW, DB order and their harmonics and combinations.
As shown in Fig.~\ref{fig:2}(c), we identify that they are moir\'e vectors originated from the combination of two independent Bragg vectors of the atomic lattice of \ce{NbSe2} and one of the Bragg vectors of graphene.
Similar moir\'e patterns originated from the higher-order combination of the lattice Bragg vectors have been observed in the \ce{MoTe2} monolayer on graphene~\cite{Pham2022}.

Since the moir\'e modulation can affect the potential landscape in a way sensitive to the twist angle, it is important to search for their signatures in the electronic states.
We focus on the quasiparticle interference (QPI) patterns that reflect momentum-space characters of quasiparticles through the scattering interference.
We found clear QPI patterns in $L(\mathbf{r},E)$ around the defects (Fig.~\ref{fig:2}(d)), which appear as a hexagonal signal in $L(\mathbf{q},E)$ with flat segments perpendicular to the $\Gamma$-M direction (Fig.~\ref{fig:2}(e)).
The $\mathbf{q}$ vector of this QPI pattern shows a hole-like energy dispersion as shown in Fig.~\ref{fig:2}(g).

The low energy band structure of \ce{NbSe2} is characterized by the hole bands centered at $\Gamma$ and K points.
Figure~\ref{fig:2}(f) shows the calculated spin-resolved spectral function based on the three-orbital tight-binding model~\cite{Liu2013}.
The $\Gamma$-centered band has a hexagonal shape in momentum space, and the scattering across the flat sides of the hexagonal spectral function (green arrow) is allowed by the spin selection rule.
This naturally explains the observed hexagonal QPI pattern in $\mathbf{q}$ space.
The same hole band also exists in bulk 2\textit{H}-\ce{NbSe2}, giving rise to similar QPI patterns~\cite{Arguello2015,Gao2018}.
Therefore, it is likely that the basic band structure is maintained even in the presence of the moir\'e potential.


Next, we describe the effect of twisted stacking on superconductivity.
As shown in Fig.~\ref{fig:3}(a), a superconducting gap opens in the low-energy tunneling spectrum.
To examine the relation between the various spatial modulations ($3 \times 3$ CDW, DB order, moir\'e, QPI, etc.) and superconductivity, we investigate the spatial variation of the gap amplitude $\Delta(\mathbf{r})$, which we phenomenologically define as the half of the energy separation between the gap-edge peaks.
(See the Supplementary Materials for details.)
We have found that $\Delta(\mathbf{r})$ is reasonably uniform with the full width at half maximum of the histogram of $\Delta(\mathbf{r})$ less than 10\% of the average gap amplitude (Fig.~\ref{fig:3}(b)). Nevertheless, a detailed examination of $\Delta(\mathbf{r})$ reveals some spatial patterns (Fig. 3C).
By taking a Fourier transformation of $\Delta(\mathbf{r})$, we have identified the $3 \times 3$ CDW and QPI modulations originating from \ce{NbSe2} monolayer but there is no DB order nor moir\'e pattern related to the underlying graphene substrate (Fig.~\ref{fig:3}(d)).
Thus, the superconducting pairing interaction is governed primarily by the \ce{NbSe2} monolayer.

It should be noted that the observed gap spectrum, shown in Fig.~\ref{fig:3}(a), is broader than expected from thermal broadening alone at the measurement temperature of \SI{90}{mK}, indicating the presence of an additional energy broadening mechanism.
We consider the energy exchange between tunneling electrons and the environment, which is known as the dynamical Coulomb blockade (DCB) effect~\cite{Brun2012,Devoret1990,Ingold1992,Ast2016}.
We evaluate the superconducting-gap spectrum in the presence of the DCB effect as shown by the red line in Fig.~\ref{fig:3}(a). (See the Supplementary Materials for details.) Although the broadened gap-edge peaks are well reproduced by the DCB model, a discrepancy appears near $V = 0$. The DCB model predicts a negligibly small weight, while an apparent residual LDOS remains in the observed spectrum.
This observation suggests that there is a gap-filling mechanism in the twisted stack of the \ce{NbSe2} monolayer on graphene.


To investigate the origin of the residual LDOS near $V = 0$, we performed high-energy-resolution SI-STM in the superconducting-gap energy scale.
Here, we analyze the raw $g(\mathbf{r}, E)$ since $L(\mathbf{r}, E)$ cannot be defined at $V = 0$ where $I = 0$. The influence of the extrinsic set-point effect will be examined later.
Figure~\ref{fig:4}(a) shows the $g(\mathbf{q},\SI{0}{meV})$ map obtained by Fourier transforming the $g(\mathbf{r},\SI{0}{meV})$ map of the twisted stack with $\theta = \ang{24}$.
Besides the hexagonal QPI signals (green segments), there are spots (yellow circles) that break the mirror symmetry with respect to the $\Gamma$-M and $\Gamma$-K lines of the \ce{NbSe2} monolayer, which we call chiral spots hereafter.

We can safely conclude that the chiral spots originate from the Bogoliubov quasiparticles in the superconducting state because they disappear in the field-induced normal state, leaving only the hexagonal QPI signals associated with the normal-state hole band (Fig.~\ref{fig:4}(b)).
Note that the set-point effect in $g(\mathbf{r}, E)$ has nothing to do with the chiral spots, since our set-point of $V_\mathrm{set} = \SI{3}{mV}$ is well above the superconducting gap, and thus the integrated LDOS in the normal and superconducting states should be the same.
The Bogoliubov-quasiparticle nature is further supported by the energy dependence of the intensity of the chiral spots.
To quantify the energy-dependent intensity of the chiral spots, we anti-symmetrize $g(\mathbf{q},E)$ with respect to the $q_y$ axis ($\Gamma$-M axis) as
\begin{math}
A(\mathbf{q},E) \equiv [g(q_x,q_y,E)-g(-q_x,q_y,E)]/[g(q_x,q_y,E)+g(-q_x,q_y,E)].
\end{math}
Note that only the signals that violate the mirror symmetry of \ce{NbSe2} remain in $A(\mathbf{q},E)$.
As shown in Fig.~\ref{fig:4} (c and d), $A(\mathbf{q},\SI{0}{meV})$ in the superconducting state shows chiral spots, which diminishes in the normal state under a magnetic field of \SI{5}{T}.
(More comprehensive $g(\mathbf{q},E)$ and $A(\mathbf{q},E)$ for the samples with different twist angles are shown in the Supplementary Materials.)
Figure~\ref{fig:4} (e and f) depict averaged tunneling spectra and $\mathbf{q}$-integrated absolute value of $A(\mathbf{q},E)$, respectively, for the two twisted stacks with $\theta = \ang{24}$ and $\theta = \ang{28}$.
The chiral signals appear inside the superconducting gap only.
We have also confirmed that the sample with $\theta = \ang{0}$, which maintains the mirror symmetry of \ce{NbSe2}, does not exhibit a chiral modulation.
(See the Supplementary Materials.)
Therefore, we conclude that the chiral spots found in $g(\mathbf{q},E)$ represent the modulations characteristic to the superconducting state of the twisted stack.


Our SI-STM experiments have revealed that the residual LDOS appears inside the superconducting gap of the twisted stacks of \ce{NbSe2} monolayer on graphene, showing the spatial modulations that do not respect the symmetry of the superconducting \ce{NbSe2} monolayer. 
The residual LDOS often appears if the superconducting gap is nodal or strongly anisotropic.
The \ce{NbSe2} monolayer on graphene breaks both in-plane and out-of-plane space inversion symmetries, allowing the parity mixing of spin-singlet and spin-triplet superconductivity~\cite{Gorkov2001}.
Therefore, in principle, the superconducting gap could have an anisotropic component.
However, as far as superconductivity occurs in the \ce{NbSe2} monolayer, the positions of nodes or gap minima, and thus the low-energy Bogoliubov quasiparticles, should respect the symmetry of \ce{NbSe2}, being inconsistent with the observations.

If the graphene layer possesses pair breaking impurities or other objects, the residual LDOS may result.
The DB between the topmost SiC layer and the buffer layer graphene possesses an unpaired electron with a spin.
Such a magnetic defect may be a candidate for the pair breaker.
However, spatial patterns related to the DB order are detected neither in the superconducting gap map (Fig.~\ref{fig:3}(c)) nor in the spatial variation of the residual LDOS (Fig.~\ref{fig:4}(a)).
Therefore, the DB is unlikely to play a significant role in the observed residual LDOS.

It is apparent that neither the \ce{NbSe2} monolayer nor the graphene layer alone can account for the observed incommensurate modulations and that the twisted stack as a whole should be considered.
We consider the proximity effect that induces superconductivity in graphene from \ce{NbSe2}.
To have a sizable proximity effect, the \ce{NbSe2} and graphene Fermi surfaces should overlap in momentum space.
As pointed out by Gani \textit{et al.}, the graphene Fermi surfaces can overlap with the $\Gamma$- and K-centered pockets of \ce{NbSe2} in the extended zone scheme for the twist angle near $\theta = \ang{30}$ and $\theta = \ang{0}$, respectively~\cite{Gani2019}. However, their approach is based on graphene Fermi surfaces, and is not suitable for the analysis of our results, which are essentially related to the property of \ce{NbSe2} Fermi surfaces. Therefore, we instead consider the position of the graphene Fermi surfaces in the first Brillouin zone of \ce{NbSe2}, following the method of Ref~\cite{Li2019}. Note that we assume the dominant contribution comes from the innermost graphene's valleys in the extended zone scheme and the effect of other outer valleys is negligible. From this approach, near $\theta = \ang{30}$, there appear a sextet of regions around the $\Gamma$-centered pocket where the Fermi surfaces of both layers overlap in the reduced zone scheme (Fig.~\ref{fig:5}(a)).
The proximity effect should induce the superconducting gap in graphene, and in turn reduce the gap of \ce{NbSe2} in the sextet regions, generating low-energy Bogoliubov quasiparticles.
In the presence of quasiparticle scatterers, we can expect Bogoliubov QPI patterns characterized by the wave vectors connecting the sextet regions (Fig.~\ref{fig:5}(a)).
This phenomenological sextet model explains all the observed chiral spots (Fig.~\ref{fig:5}(b)) and their twist angle dependence (Fig.~\ref{fig:5}(c)).

Overall, our observations indicate that twisted stacking helps to induce emergent superconducting phenomena that are absent in the original superconducting layer.
Future technical advances in sample preparation may allow us to systematically control the twist angle, leading to a deeper understanding of how the coupling between \ce{NbSe2} and graphene affects superconductivity.
While the present work has focused on the \ce{NbSe2} monolayer on graphene, the insights gained here should also be relevant to other stacked two-dimensional superconductors containing other transition metal dichalcogenides and twisted bilayer graphene.
In addition to bilayer stacking, chiral stacking of multilayers consisting of different types of atomic layers may harbor yet unknown superconducting and topological properties, leading to superconducting twistronics.

\bibliography{NbSe2_twist}

\begin{thebibliography}{48}%
\makeatletter
\providecommand \@ifxundefined [1]{%
 \@ifx{#1\undefined}
}%
\providecommand \@ifnum [1]{%
 \ifnum #1\expandafter \@firstoftwo
 \else \expandafter \@secondoftwo
 \fi
}%
\providecommand \@ifx [1]{%
 \ifx #1\expandafter \@firstoftwo
 \else \expandafter \@secondoftwo
 \fi
}%
\providecommand \natexlab [1]{#1}%
\providecommand \enquote  [1]{``#1''}%
\providecommand \bibnamefont  [1]{#1}%
\providecommand \bibfnamefont [1]{#1}%
\providecommand \citenamefont [1]{#1}%
\providecommand \href@noop [0]{\@secondoftwo}%
\providecommand \href [0]{\begingroup \@sanitize@url \@href}%
\providecommand \@href[1]{\@@startlink{#1}\@@href}%
\providecommand \@@href[1]{\endgroup#1\@@endlink}%
\providecommand \@sanitize@url [0]{\catcode `\\12\catcode `\$12\catcode `\&12\catcode `\#12\catcode `\^12\catcode `\_12\catcode `\%12\relax}%
\providecommand \@@startlink[1]{}%
\providecommand \@@endlink[0]{}%
\providecommand \url  [0]{\begingroup\@sanitize@url \@url }%
\providecommand \@url [1]{\endgroup\@href {#1}{\urlprefix }}%
\providecommand \urlprefix  [0]{URL }%
\providecommand \Eprint [0]{\href }%
\providecommand \doibase [0]{https://doi.org/}%
\providecommand \selectlanguage [0]{\@gobble}%
\providecommand \bibinfo  [0]{\@secondoftwo}%
\providecommand \bibfield  [0]{\@secondoftwo}%
\providecommand \translation [1]{[#1]}%
\providecommand \BibitemOpen [0]{}%
\providecommand \bibitemStop [0]{}%
\providecommand \bibitemNoStop [0]{.\EOS\space}%
\providecommand \EOS [0]{\spacefactor3000\relax}%
\providecommand \BibitemShut  [1]{\csname bibitem#1\endcsname}%
\let\auto@bib@innerbib\@empty
\bibitem [{\citenamefont {Geim}\ and\ \citenamefont {Grigorieva}(2013)}]{Geim2013}%
  \BibitemOpen
  \bibfield  {author} {\bibinfo {author} {\bibfnamefont {A.~K.}\ \bibnamefont {Geim}}\ and\ \bibinfo {author} {\bibfnamefont {I.~V.}\ \bibnamefont {Grigorieva}},\ }\bibfield  {title} {\bibinfo {title} {Van der waals heterostructures},\ }\href {https://doi.org/10.1038/nature12385} {\bibfield  {journal} {\bibinfo  {journal} {Nature}\ }\textbf {\bibinfo {volume} {499}},\ \bibinfo {pages} {419} (\bibinfo {year} {2013})}\BibitemShut {NoStop}%
\bibitem [{\citenamefont {Novoselov}\ \emph {et~al.}(2016)\citenamefont {Novoselov}, \citenamefont {Mishchenko}, \citenamefont {Carvalho},\ and\ \citenamefont {Neto}}]{Novoselov2016}%
  \BibitemOpen
  \bibfield  {author} {\bibinfo {author} {\bibfnamefont {K.~S.}\ \bibnamefont {Novoselov}}, \bibinfo {author} {\bibfnamefont {A.}~\bibnamefont {Mishchenko}}, \bibinfo {author} {\bibfnamefont {A.}~\bibnamefont {Carvalho}},\ and\ \bibinfo {author} {\bibfnamefont {A.~H.~C.}\ \bibnamefont {Neto}},\ }\bibfield  {title} {\bibinfo {title} {{2D} materials and van der waals heterostructures},\ }\href {https://doi.org/10.1126/science.aac9439} {\bibfield  {journal} {\bibinfo  {journal} {Science}\ }\textbf {\bibinfo {volume} {353}},\ \bibinfo {pages} {aac9439} (\bibinfo {year} {2016})}\BibitemShut {NoStop}%
\bibitem [{\citenamefont {Liu}\ \emph {et~al.}(2019)\citenamefont {Liu}, \citenamefont {Huang},\ and\ \citenamefont {Duan}}]{Liu2019}%
  \BibitemOpen
  \bibfield  {author} {\bibinfo {author} {\bibfnamefont {Y.}~\bibnamefont {Liu}}, \bibinfo {author} {\bibfnamefont {Y.}~\bibnamefont {Huang}},\ and\ \bibinfo {author} {\bibfnamefont {X.}~\bibnamefont {Duan}},\ }\bibfield  {title} {\bibinfo {title} {Van der waals integration before and beyond two-dimensional materials},\ }\href {https://doi.org/10.1038/s41586-019-1013-x} {\bibfield  {journal} {\bibinfo  {journal} {Nature}\ }\textbf {\bibinfo {volume} {567}},\ \bibinfo {pages} {323} (\bibinfo {year} {2019})}\BibitemShut {NoStop}%
\bibitem [{\citenamefont {Novoselov}\ \emph {et~al.}(2004)\citenamefont {Novoselov}, \citenamefont {Geim}, \citenamefont {Morozov}, \citenamefont {Jiang}, \citenamefont {Zhang}, \citenamefont {Dubonos}, \citenamefont {Grigorieva},\ and\ \citenamefont {Firsov}}]{Novoselov2004}%
  \BibitemOpen
  \bibfield  {author} {\bibinfo {author} {\bibfnamefont {K.~S.}\ \bibnamefont {Novoselov}}, \bibinfo {author} {\bibfnamefont {A.~K.}\ \bibnamefont {Geim}}, \bibinfo {author} {\bibfnamefont {S.~V.}\ \bibnamefont {Morozov}}, \bibinfo {author} {\bibfnamefont {D.}~\bibnamefont {Jiang}}, \bibinfo {author} {\bibfnamefont {Y.}~\bibnamefont {Zhang}}, \bibinfo {author} {\bibfnamefont {S.~V.}\ \bibnamefont {Dubonos}}, \bibinfo {author} {\bibfnamefont {I.~V.}\ \bibnamefont {Grigorieva}},\ and\ \bibinfo {author} {\bibfnamefont {A.~A.}\ \bibnamefont {Firsov}},\ }\bibfield  {title} {\bibinfo {title} {Electric field effect in atomically thin carbon films},\ }\href {https://doi.org/10.1126/science.1102896} {\bibfield  {journal} {\bibinfo  {journal} {Science}\ }\textbf {\bibinfo {volume} {306}},\ \bibinfo {pages} {666} (\bibinfo {year} {2004})}\BibitemShut {NoStop}%
\bibitem [{\citenamefont {Mak}\ \emph {et~al.}(2010)\citenamefont {Mak}, \citenamefont {Lee}, \citenamefont {Hone}, \citenamefont {Shan},\ and\ \citenamefont {Heinz}}]{Mak2010}%
  \BibitemOpen
  \bibfield  {author} {\bibinfo {author} {\bibfnamefont {K.~F.}\ \bibnamefont {Mak}}, \bibinfo {author} {\bibfnamefont {C.}~\bibnamefont {Lee}}, \bibinfo {author} {\bibfnamefont {J.}~\bibnamefont {Hone}}, \bibinfo {author} {\bibfnamefont {J.}~\bibnamefont {Shan}},\ and\ \bibinfo {author} {\bibfnamefont {T.~F.}\ \bibnamefont {Heinz}},\ }\bibfield  {title} {\bibinfo {title} {Atomically thin {${\mathrm{MoS}}_{2}$}: A new direct-gap semiconductor},\ }\href {https://doi.org/10.1103/PhysRevLett.105.136805} {\bibfield  {journal} {\bibinfo  {journal} {Phys. Rev. Lett.}\ }\textbf {\bibinfo {volume} {105}},\ \bibinfo {pages} {136805} (\bibinfo {year} {2010})}\BibitemShut {NoStop}%
\bibitem [{\citenamefont {Splendiani}\ \emph {et~al.}(2010)\citenamefont {Splendiani}, \citenamefont {Sun}, \citenamefont {Zhang}, \citenamefont {Li}, \citenamefont {Kim}, \citenamefont {Chim}, \citenamefont {Galli},\ and\ \citenamefont {Wang}}]{Splendiani2010}%
  \BibitemOpen
  \bibfield  {author} {\bibinfo {author} {\bibfnamefont {A.}~\bibnamefont {Splendiani}}, \bibinfo {author} {\bibfnamefont {L.}~\bibnamefont {Sun}}, \bibinfo {author} {\bibfnamefont {Y.}~\bibnamefont {Zhang}}, \bibinfo {author} {\bibfnamefont {T.}~\bibnamefont {Li}}, \bibinfo {author} {\bibfnamefont {J.}~\bibnamefont {Kim}}, \bibinfo {author} {\bibfnamefont {C.-Y.}\ \bibnamefont {Chim}}, \bibinfo {author} {\bibfnamefont {G.}~\bibnamefont {Galli}},\ and\ \bibinfo {author} {\bibfnamefont {F.}~\bibnamefont {Wang}},\ }\bibfield  {title} {\bibinfo {title} {Emerging photoluminescence in monolayer {$\mathrm{MoS}_{2}$}},\ }\href {https://doi.org/10.1021/nl903868w} {\bibfield  {journal} {\bibinfo  {journal} {Nano Letters}\ }\textbf {\bibinfo {volume} {10}},\ \bibinfo {pages} {1271} (\bibinfo {year} {2010})}\BibitemShut {NoStop}%
\bibitem [{\citenamefont {Deng}\ \emph {et~al.}(2018)\citenamefont {Deng}, \citenamefont {Yu}, \citenamefont {Song}, \citenamefont {Zhang}, \citenamefont {Wang}, \citenamefont {Sun}, \citenamefont {Yi}, \citenamefont {Wu}, \citenamefont {Wu}, \citenamefont {Zhu}, \citenamefont {Wang}, \citenamefont {Chen},\ and\ \citenamefont {Zhang}}]{Deng2018}%
  \BibitemOpen
  \bibfield  {author} {\bibinfo {author} {\bibfnamefont {Y.}~\bibnamefont {Deng}}, \bibinfo {author} {\bibfnamefont {Y.}~\bibnamefont {Yu}}, \bibinfo {author} {\bibfnamefont {Y.}~\bibnamefont {Song}}, \bibinfo {author} {\bibfnamefont {J.}~\bibnamefont {Zhang}}, \bibinfo {author} {\bibfnamefont {N.~Z.}\ \bibnamefont {Wang}}, \bibinfo {author} {\bibfnamefont {Z.}~\bibnamefont {Sun}}, \bibinfo {author} {\bibfnamefont {Y.}~\bibnamefont {Yi}}, \bibinfo {author} {\bibfnamefont {Y.~Z.}\ \bibnamefont {Wu}}, \bibinfo {author} {\bibfnamefont {S.}~\bibnamefont {Wu}}, \bibinfo {author} {\bibfnamefont {J.}~\bibnamefont {Zhu}}, \bibinfo {author} {\bibfnamefont {J.}~\bibnamefont {Wang}}, \bibinfo {author} {\bibfnamefont {X.~H.}\ \bibnamefont {Chen}},\ and\ \bibinfo {author} {\bibfnamefont {Y.}~\bibnamefont {Zhang}},\ }\bibfield  {title} {\bibinfo {title} {Gate-tunable room-temperature ferromagnetism in two-dimensional {$\mathrm{Fe}_{3}\mathrm{GeTe}_{2}$}},\ }\href {https://doi.org/10.1038/s41586-018-0626-9} {\bibfield
  {journal} {\bibinfo  {journal} {Nature}\ }\textbf {\bibinfo {volume} {563}},\ \bibinfo {pages} {94} (\bibinfo {year} {2018})}\BibitemShut {NoStop}%
\bibitem [{\citenamefont {Seyler}\ \emph {et~al.}(2018)\citenamefont {Seyler}, \citenamefont {Zhong}, \citenamefont {Klein}, \citenamefont {Gao}, \citenamefont {Zhang}, \citenamefont {Huang}, \citenamefont {Navarro-Moratalla}, \citenamefont {Yang}, \citenamefont {Cobden}, \citenamefont {McGuire}, \citenamefont {Yao}, \citenamefont {Xiao}, \citenamefont {Jarillo-Herrero},\ and\ \citenamefont {Xu}}]{Seyler2018}%
  \BibitemOpen
  \bibfield  {author} {\bibinfo {author} {\bibfnamefont {K.~L.}\ \bibnamefont {Seyler}}, \bibinfo {author} {\bibfnamefont {D.}~\bibnamefont {Zhong}}, \bibinfo {author} {\bibfnamefont {D.~R.}\ \bibnamefont {Klein}}, \bibinfo {author} {\bibfnamefont {S.}~\bibnamefont {Gao}}, \bibinfo {author} {\bibfnamefont {X.}~\bibnamefont {Zhang}}, \bibinfo {author} {\bibfnamefont {B.}~\bibnamefont {Huang}}, \bibinfo {author} {\bibfnamefont {E.}~\bibnamefont {Navarro-Moratalla}}, \bibinfo {author} {\bibfnamefont {L.}~\bibnamefont {Yang}}, \bibinfo {author} {\bibfnamefont {D.~H.}\ \bibnamefont {Cobden}}, \bibinfo {author} {\bibfnamefont {M.~A.}\ \bibnamefont {McGuire}}, \bibinfo {author} {\bibfnamefont {W.}~\bibnamefont {Yao}}, \bibinfo {author} {\bibfnamefont {D.}~\bibnamefont {Xiao}}, \bibinfo {author} {\bibfnamefont {P.}~\bibnamefont {Jarillo-Herrero}},\ and\ \bibinfo {author} {\bibfnamefont {X.}~\bibnamefont {Xu}},\ }\bibfield  {title} {\bibinfo {title} {Ligand-field helical luminescence in a {2D} ferromagnetic
  insulator},\ }\href {https://doi.org/10.1038/s41567-017-0006-7} {\bibfield  {journal} {\bibinfo  {journal} {Nature Physics}\ }\textbf {\bibinfo {volume} {14}},\ \bibinfo {pages} {277} (\bibinfo {year} {2018})}\BibitemShut {NoStop}%
\bibitem [{\citenamefont {Fu}\ and\ \citenamefont {Kane}(2008)}]{Fu2008}%
  \BibitemOpen
  \bibfield  {author} {\bibinfo {author} {\bibfnamefont {L.}~\bibnamefont {Fu}}\ and\ \bibinfo {author} {\bibfnamefont {C.~L.}\ \bibnamefont {Kane}},\ }\bibfield  {title} {\bibinfo {title} {Superconducting proximity effect and majorana fermions at the surface of a topological insulator},\ }\href {https://doi.org/10.1103/PhysRevLett.100.096407} {\bibfield  {journal} {\bibinfo  {journal} {Phys. Rev. Lett.}\ }\textbf {\bibinfo {volume} {100}},\ \bibinfo {pages} {096407} (\bibinfo {year} {2008})}\BibitemShut {NoStop}%
\bibitem [{\citenamefont {Xu}\ \emph {et~al.}(2014)\citenamefont {Xu}, \citenamefont {Liu}, \citenamefont {Wang}, \citenamefont {Ge}, \citenamefont {Liu}, \citenamefont {Yang}, \citenamefont {Chen}, \citenamefont {Liu}, \citenamefont {Xu}, \citenamefont {Gao}, \citenamefont {Qian}, \citenamefont {Zhang},\ and\ \citenamefont {Jia}}]{Xu2014}%
  \BibitemOpen
  \bibfield  {author} {\bibinfo {author} {\bibfnamefont {J.-P.}\ \bibnamefont {Xu}}, \bibinfo {author} {\bibfnamefont {C.}~\bibnamefont {Liu}}, \bibinfo {author} {\bibfnamefont {M.-X.}\ \bibnamefont {Wang}}, \bibinfo {author} {\bibfnamefont {J.}~\bibnamefont {Ge}}, \bibinfo {author} {\bibfnamefont {Z.-L.}\ \bibnamefont {Liu}}, \bibinfo {author} {\bibfnamefont {X.}~\bibnamefont {Yang}}, \bibinfo {author} {\bibfnamefont {Y.}~\bibnamefont {Chen}}, \bibinfo {author} {\bibfnamefont {Y.}~\bibnamefont {Liu}}, \bibinfo {author} {\bibfnamefont {Z.-A.}\ \bibnamefont {Xu}}, \bibinfo {author} {\bibfnamefont {C.-L.}\ \bibnamefont {Gao}}, \bibinfo {author} {\bibfnamefont {D.}~\bibnamefont {Qian}}, \bibinfo {author} {\bibfnamefont {F.-C.}\ \bibnamefont {Zhang}},\ and\ \bibinfo {author} {\bibfnamefont {J.-F.}\ \bibnamefont {Jia}},\ }\bibfield  {title} {\bibinfo {title} {Artificial topological superconductor by the proximity effect},\ }\href {https://doi.org/10.1103/PhysRevLett.112.217001} {\bibfield  {journal} {\bibinfo
  {journal} {Phys. Rev. Lett.}\ }\textbf {\bibinfo {volume} {112}},\ \bibinfo {pages} {217001} (\bibinfo {year} {2014})}\BibitemShut {NoStop}%
\bibitem [{\citenamefont {Xu}\ \emph {et~al.}(2015)\citenamefont {Xu}, \citenamefont {Wang}, \citenamefont {Liu}, \citenamefont {Ge}, \citenamefont {Yang}, \citenamefont {Liu}, \citenamefont {Xu}, \citenamefont {Guan}, \citenamefont {Gao}, \citenamefont {Qian}, \citenamefont {Liu}, \citenamefont {Wang}, \citenamefont {Zhang}, \citenamefont {Xue},\ and\ \citenamefont {Jia}}]{Xu2015}%
  \BibitemOpen
  \bibfield  {author} {\bibinfo {author} {\bibfnamefont {J.-P.}\ \bibnamefont {Xu}}, \bibinfo {author} {\bibfnamefont {M.-X.}\ \bibnamefont {Wang}}, \bibinfo {author} {\bibfnamefont {Z.~L.}\ \bibnamefont {Liu}}, \bibinfo {author} {\bibfnamefont {J.-F.}\ \bibnamefont {Ge}}, \bibinfo {author} {\bibfnamefont {X.}~\bibnamefont {Yang}}, \bibinfo {author} {\bibfnamefont {C.}~\bibnamefont {Liu}}, \bibinfo {author} {\bibfnamefont {Z.~A.}\ \bibnamefont {Xu}}, \bibinfo {author} {\bibfnamefont {D.}~\bibnamefont {Guan}}, \bibinfo {author} {\bibfnamefont {C.~L.}\ \bibnamefont {Gao}}, \bibinfo {author} {\bibfnamefont {D.}~\bibnamefont {Qian}}, \bibinfo {author} {\bibfnamefont {Y.}~\bibnamefont {Liu}}, \bibinfo {author} {\bibfnamefont {Q.-H.}\ \bibnamefont {Wang}}, \bibinfo {author} {\bibfnamefont {F.-C.}\ \bibnamefont {Zhang}}, \bibinfo {author} {\bibfnamefont {Q.-K.}\ \bibnamefont {Xue}},\ and\ \bibinfo {author} {\bibfnamefont {J.-F.}\ \bibnamefont {Jia}},\ }\bibfield  {title} {\bibinfo {title} {Experimental detection of
  a majorana mode in the core of a magnetic vortex inside a topological insulator-superconductor {${\mathrm{Bi}}_{2}{\mathrm{Te}}_{3}/{\mathrm{NbSe}}_{2}$} heterostructure},\ }\href {https://doi.org/10.1103/PhysRevLett.114.017001} {\bibfield  {journal} {\bibinfo  {journal} {Phys. Rev. Lett.}\ }\textbf {\bibinfo {volume} {114}},\ \bibinfo {pages} {017001} (\bibinfo {year} {2015})}\BibitemShut {NoStop}%
\bibitem [{\citenamefont {L{\"u}pke}\ \emph {et~al.}(2020)\citenamefont {L{\"u}pke}, \citenamefont {Waters}, \citenamefont {de~la Barrera}, \citenamefont {Widom}, \citenamefont {Mandrus}, \citenamefont {Yan}, \citenamefont {Feenstra},\ and\ \citenamefont {Hunt}}]{Lupke2020}%
  \BibitemOpen
  \bibfield  {author} {\bibinfo {author} {\bibfnamefont {F.}~\bibnamefont {L{\"u}pke}}, \bibinfo {author} {\bibfnamefont {D.}~\bibnamefont {Waters}}, \bibinfo {author} {\bibfnamefont {S.~C.}\ \bibnamefont {de~la Barrera}}, \bibinfo {author} {\bibfnamefont {M.}~\bibnamefont {Widom}}, \bibinfo {author} {\bibfnamefont {D.~G.}\ \bibnamefont {Mandrus}}, \bibinfo {author} {\bibfnamefont {J.}~\bibnamefont {Yan}}, \bibinfo {author} {\bibfnamefont {R.~M.}\ \bibnamefont {Feenstra}},\ and\ \bibinfo {author} {\bibfnamefont {B.~M.}\ \bibnamefont {Hunt}},\ }\bibfield  {title} {\bibinfo {title} {Proximity-induced superconducting gap in the quantum spin hall edge state of monolayer {$\mathrm{WTe}_{2}$}},\ }\href {https://doi.org/10.1038/s41567-020-0816-x} {\bibfield  {journal} {\bibinfo  {journal} {Nature Physics}\ }\textbf {\bibinfo {volume} {16}},\ \bibinfo {pages} {526} (\bibinfo {year} {2020})}\BibitemShut {NoStop}%
\bibitem [{\citenamefont {Kezilebieke}\ \emph {et~al.}(2022)\citenamefont {Kezilebieke}, \citenamefont {Vaňo}, \citenamefont {Huda}, \citenamefont {Aapro}, \citenamefont {Ganguli}, \citenamefont {Liljeroth},\ and\ \citenamefont {Lado}}]{Kezilebieke2022}%
  \BibitemOpen
  \bibfield  {author} {\bibinfo {author} {\bibfnamefont {S.}~\bibnamefont {Kezilebieke}}, \bibinfo {author} {\bibfnamefont {V.}~\bibnamefont {Vaňo}}, \bibinfo {author} {\bibfnamefont {M.~N.}\ \bibnamefont {Huda}}, \bibinfo {author} {\bibfnamefont {M.}~\bibnamefont {Aapro}}, \bibinfo {author} {\bibfnamefont {S.~C.}\ \bibnamefont {Ganguli}}, \bibinfo {author} {\bibfnamefont {P.}~\bibnamefont {Liljeroth}},\ and\ \bibinfo {author} {\bibfnamefont {J.~L.}\ \bibnamefont {Lado}},\ }\bibfield  {title} {\bibinfo {title} {Moir{\'e}-enabled topological superconductivity},\ }\href {https://doi.org/10.1021/acs.nanolett.1c03856} {\bibfield  {journal} {\bibinfo  {journal} {Nano Letters}\ }\textbf {\bibinfo {volume} {22}},\ \bibinfo {pages} {328} (\bibinfo {year} {2022})}\BibitemShut {NoStop}%
\bibitem [{\citenamefont {Bistritzer}\ and\ \citenamefont {MacDonald}(2011)}]{Bistritzer2011}%
  \BibitemOpen
  \bibfield  {author} {\bibinfo {author} {\bibfnamefont {R.}~\bibnamefont {Bistritzer}}\ and\ \bibinfo {author} {\bibfnamefont {A.~H.}\ \bibnamefont {MacDonald}},\ }\bibfield  {title} {\bibinfo {title} {Moir{\'e} bands in twisted double-layer graphene},\ }\href {https://doi.org/10.1073/pnas.1108174108} {\bibfield  {journal} {\bibinfo  {journal} {Proceedings of the National Academy of Sciences}\ }\textbf {\bibinfo {volume} {108}},\ \bibinfo {pages} {12233} (\bibinfo {year} {2011})}\BibitemShut {NoStop}%
\bibitem [{\citenamefont {Kennes}\ \emph {et~al.}(2021)\citenamefont {Kennes}, \citenamefont {Claassen}, \citenamefont {Xian}, \citenamefont {Georges}, \citenamefont {Millis}, \citenamefont {Hone}, \citenamefont {Dean}, \citenamefont {Basov}, \citenamefont {Pasupathy},\ and\ \citenamefont {Rubio}}]{Kennes2021}%
  \BibitemOpen
  \bibfield  {author} {\bibinfo {author} {\bibfnamefont {D.~M.}\ \bibnamefont {Kennes}}, \bibinfo {author} {\bibfnamefont {M.}~\bibnamefont {Claassen}}, \bibinfo {author} {\bibfnamefont {L.}~\bibnamefont {Xian}}, \bibinfo {author} {\bibfnamefont {A.}~\bibnamefont {Georges}}, \bibinfo {author} {\bibfnamefont {A.~J.}\ \bibnamefont {Millis}}, \bibinfo {author} {\bibfnamefont {J.}~\bibnamefont {Hone}}, \bibinfo {author} {\bibfnamefont {C.~R.}\ \bibnamefont {Dean}}, \bibinfo {author} {\bibfnamefont {D.~N.}\ \bibnamefont {Basov}}, \bibinfo {author} {\bibfnamefont {A.~N.}\ \bibnamefont {Pasupathy}},\ and\ \bibinfo {author} {\bibfnamefont {A.}~\bibnamefont {Rubio}},\ }\bibfield  {title} {\bibinfo {title} {Moir{\'e} heterostructures as a condensed-matter quantum simulator},\ }\href {https://doi.org/10.1038/s41567-020-01154-3} {\bibfield  {journal} {\bibinfo  {journal} {Nature Physics}\ }\textbf {\bibinfo {volume} {17}},\ \bibinfo {pages} {155} (\bibinfo {year} {2021})}\BibitemShut {NoStop}%
\bibitem [{\citenamefont {Cao}\ \emph {et~al.}(2018)\citenamefont {Cao}, \citenamefont {Fatemi}, \citenamefont {Fang}, \citenamefont {Watanabe}, \citenamefont {Taniguchi}, \citenamefont {Kaxiras},\ and\ \citenamefont {Jarillo-Herrero}}]{Cao2018_1}%
  \BibitemOpen
  \bibfield  {author} {\bibinfo {author} {\bibfnamefont {Y.}~\bibnamefont {Cao}}, \bibinfo {author} {\bibfnamefont {V.}~\bibnamefont {Fatemi}}, \bibinfo {author} {\bibfnamefont {S.}~\bibnamefont {Fang}}, \bibinfo {author} {\bibfnamefont {K.}~\bibnamefont {Watanabe}}, \bibinfo {author} {\bibfnamefont {T.}~\bibnamefont {Taniguchi}}, \bibinfo {author} {\bibfnamefont {E.}~\bibnamefont {Kaxiras}},\ and\ \bibinfo {author} {\bibfnamefont {P.}~\bibnamefont {Jarillo-Herrero}},\ }\bibfield  {title} {\bibinfo {title} {Unconventional superconductivity in magic-angle graphene superlattices},\ }\href {https://doi.org/10.1038/nature26160} {\bibfield  {journal} {\bibinfo  {journal} {Nature}\ }\textbf {\bibinfo {volume} {556}},\ \bibinfo {pages} {43} (\bibinfo {year} {2018})}\BibitemShut {NoStop}%
\bibitem [{\citenamefont {Yankowitz}\ \emph {et~al.}(2019)\citenamefont {Yankowitz}, \citenamefont {Chen}, \citenamefont {Polshyn}, \citenamefont {Zhang}, \citenamefont {Watanabe}, \citenamefont {Taniguchi}, \citenamefont {Graf}, \citenamefont {Young},\ and\ \citenamefont {Dean}}]{Yankowitz2019}%
  \BibitemOpen
  \bibfield  {author} {\bibinfo {author} {\bibfnamefont {M.}~\bibnamefont {Yankowitz}}, \bibinfo {author} {\bibfnamefont {S.}~\bibnamefont {Chen}}, \bibinfo {author} {\bibfnamefont {H.}~\bibnamefont {Polshyn}}, \bibinfo {author} {\bibfnamefont {Y.}~\bibnamefont {Zhang}}, \bibinfo {author} {\bibfnamefont {K.}~\bibnamefont {Watanabe}}, \bibinfo {author} {\bibfnamefont {T.}~\bibnamefont {Taniguchi}}, \bibinfo {author} {\bibfnamefont {D.}~\bibnamefont {Graf}}, \bibinfo {author} {\bibfnamefont {A.~F.}\ \bibnamefont {Young}},\ and\ \bibinfo {author} {\bibfnamefont {C.~R.}\ \bibnamefont {Dean}},\ }\bibfield  {title} {\bibinfo {title} {Tuning superconductivity in twisted bilayer graphene},\ }\href {https://doi.org/10.1126/science.aav1910} {\bibfield  {journal} {\bibinfo  {journal} {Science}\ }\textbf {\bibinfo {volume} {363}},\ \bibinfo {pages} {1059} (\bibinfo {year} {2019})}\BibitemShut {NoStop}%
\bibitem [{\citenamefont {Lu}\ \emph {et~al.}(2019)\citenamefont {Lu}, \citenamefont {Stepanov}, \citenamefont {Yang}, \citenamefont {Xie}, \citenamefont {Aamir}, \citenamefont {Das}, \citenamefont {Urgell}, \citenamefont {Watanabe}, \citenamefont {Taniguchi}, \citenamefont {Zhang}, \citenamefont {Bachtold}, \citenamefont {MacDonald},\ and\ \citenamefont {Efetov}}]{Lu2019}%
  \BibitemOpen
  \bibfield  {author} {\bibinfo {author} {\bibfnamefont {X.}~\bibnamefont {Lu}}, \bibinfo {author} {\bibfnamefont {P.}~\bibnamefont {Stepanov}}, \bibinfo {author} {\bibfnamefont {W.}~\bibnamefont {Yang}}, \bibinfo {author} {\bibfnamefont {M.}~\bibnamefont {Xie}}, \bibinfo {author} {\bibfnamefont {M.~A.}\ \bibnamefont {Aamir}}, \bibinfo {author} {\bibfnamefont {I.}~\bibnamefont {Das}}, \bibinfo {author} {\bibfnamefont {C.}~\bibnamefont {Urgell}}, \bibinfo {author} {\bibfnamefont {K.}~\bibnamefont {Watanabe}}, \bibinfo {author} {\bibfnamefont {T.}~\bibnamefont {Taniguchi}}, \bibinfo {author} {\bibfnamefont {G.}~\bibnamefont {Zhang}}, \bibinfo {author} {\bibfnamefont {A.}~\bibnamefont {Bachtold}}, \bibinfo {author} {\bibfnamefont {A.~H.}\ \bibnamefont {MacDonald}},\ and\ \bibinfo {author} {\bibfnamefont {D.~K.}\ \bibnamefont {Efetov}},\ }\bibfield  {title} {\bibinfo {title} {Superconductors, orbital magnets and correlated states in magic-angle bilayer graphene},\ }\href
  {https://doi.org/10.1038/s41586-019-1695-0} {\bibfield  {journal} {\bibinfo  {journal} {Nature}\ }\textbf {\bibinfo {volume} {574}},\ \bibinfo {pages} {653} (\bibinfo {year} {2019})}\BibitemShut {NoStop}%
\bibitem [{\citenamefont {Stepanov}\ \emph {et~al.}(2020)\citenamefont {Stepanov}, \citenamefont {Das}, \citenamefont {Lu}, \citenamefont {Fahimniya}, \citenamefont {Watanabe}, \citenamefont {Taniguchi}, \citenamefont {Koppens}, \citenamefont {Lischner}, \citenamefont {Levitov},\ and\ \citenamefont {Efetov}}]{Stepanov2020}%
  \BibitemOpen
  \bibfield  {author} {\bibinfo {author} {\bibfnamefont {P.}~\bibnamefont {Stepanov}}, \bibinfo {author} {\bibfnamefont {I.}~\bibnamefont {Das}}, \bibinfo {author} {\bibfnamefont {X.}~\bibnamefont {Lu}}, \bibinfo {author} {\bibfnamefont {A.}~\bibnamefont {Fahimniya}}, \bibinfo {author} {\bibfnamefont {K.}~\bibnamefont {Watanabe}}, \bibinfo {author} {\bibfnamefont {T.}~\bibnamefont {Taniguchi}}, \bibinfo {author} {\bibfnamefont {F.~H.~L.}\ \bibnamefont {Koppens}}, \bibinfo {author} {\bibfnamefont {J.}~\bibnamefont {Lischner}}, \bibinfo {author} {\bibfnamefont {L.}~\bibnamefont {Levitov}},\ and\ \bibinfo {author} {\bibfnamefont {D.~K.}\ \bibnamefont {Efetov}},\ }\bibfield  {title} {\bibinfo {title} {Untying the insulating and superconducting orders in magic-angle graphene},\ }\href {https://doi.org/10.1038/s41586-020-2459-6} {\bibfield  {journal} {\bibinfo  {journal} {Nature}\ }\textbf {\bibinfo {volume} {583}},\ \bibinfo {pages} {375} (\bibinfo {year} {2020})}\BibitemShut {NoStop}%
\bibitem [{\citenamefont {Oh}\ \emph {et~al.}(2021)\citenamefont {Oh}, \citenamefont {Nuckolls}, \citenamefont {Wong}, \citenamefont {Lee}, \citenamefont {Liu}, \citenamefont {Watanabe}, \citenamefont {Taniguchi},\ and\ \citenamefont {Yazdani}}]{Oh2021}%
  \BibitemOpen
  \bibfield  {author} {\bibinfo {author} {\bibfnamefont {M.}~\bibnamefont {Oh}}, \bibinfo {author} {\bibfnamefont {K.~P.}\ \bibnamefont {Nuckolls}}, \bibinfo {author} {\bibfnamefont {D.}~\bibnamefont {Wong}}, \bibinfo {author} {\bibfnamefont {R.~L.}\ \bibnamefont {Lee}}, \bibinfo {author} {\bibfnamefont {X.}~\bibnamefont {Liu}}, \bibinfo {author} {\bibfnamefont {K.}~\bibnamefont {Watanabe}}, \bibinfo {author} {\bibfnamefont {T.}~\bibnamefont {Taniguchi}},\ and\ \bibinfo {author} {\bibfnamefont {A.}~\bibnamefont {Yazdani}},\ }\bibfield  {title} {\bibinfo {title} {Evidence for unconventional superconductivity in twisted bilayer graphene},\ }\href {https://doi.org/10.1038/s41586-021-04121-x} {\bibfield  {journal} {\bibinfo  {journal} {Nature}\ }\textbf {\bibinfo {volume} {600}},\ \bibinfo {pages} {240} (\bibinfo {year} {2021})}\BibitemShut {NoStop}%
\bibitem [{\citenamefont {Wilson}\ \emph {et~al.}(2001)\citenamefont {Wilson}, \citenamefont {Salvo},\ and\ \citenamefont {Mahajan}}]{Wilson2001}%
  \BibitemOpen
  \bibfield  {author} {\bibinfo {author} {\bibfnamefont {J.~A.}\ \bibnamefont {Wilson}}, \bibinfo {author} {\bibfnamefont {F.~J.~D.}\ \bibnamefont {Salvo}},\ and\ \bibinfo {author} {\bibfnamefont {S.}~\bibnamefont {Mahajan}},\ }\bibfield  {title} {\bibinfo {title} {Charge-density waves and superlattices in the metallic layered transition metal dichalcogenides},\ }\href {https://doi.org/10.1080/00018730110102718} {\bibfield  {journal} {\bibinfo  {journal} {Advances in Physics}\ }\textbf {\bibinfo {volume} {50}},\ \bibinfo {pages} {1171} (\bibinfo {year} {2001})}\BibitemShut {NoStop}%
\bibitem [{\citenamefont {Moncton}\ \emph {et~al.}(1975)\citenamefont {Moncton}, \citenamefont {Axe},\ and\ \citenamefont {DiSalvo}}]{Moncton1975}%
  \BibitemOpen
  \bibfield  {author} {\bibinfo {author} {\bibfnamefont {D.~E.}\ \bibnamefont {Moncton}}, \bibinfo {author} {\bibfnamefont {J.~D.}\ \bibnamefont {Axe}},\ and\ \bibinfo {author} {\bibfnamefont {F.~J.}\ \bibnamefont {DiSalvo}},\ }\bibfield  {title} {\bibinfo {title} {Study of superlattice formation in {$2H$-Nb${\mathrm{Se}}_{2}$} and {$2H$-Ta${\mathrm{Se}}_{2}$} by neutron scattering},\ }\href {https://doi.org/10.1103/PhysRevLett.34.734} {\bibfield  {journal} {\bibinfo  {journal} {Phys. Rev. Lett.}\ }\textbf {\bibinfo {volume} {34}},\ \bibinfo {pages} {734} (\bibinfo {year} {1975})}\BibitemShut {NoStop}%
\bibitem [{\citenamefont {Revolinsky}\ \emph {et~al.}(1965)\citenamefont {Revolinsky}, \citenamefont {Spiering},\ and\ \citenamefont {Beerntsen}}]{Revolinsky1965}%
  \BibitemOpen
  \bibfield  {author} {\bibinfo {author} {\bibfnamefont {E.}~\bibnamefont {Revolinsky}}, \bibinfo {author} {\bibfnamefont {G.}~\bibnamefont {Spiering}},\ and\ \bibinfo {author} {\bibfnamefont {D.}~\bibnamefont {Beerntsen}},\ }\bibfield  {title} {\bibinfo {title} {Superconductivity in the niobium-selenium system},\ }\href {https://doi.org/https://doi.org/10.1016/0022-3697(65)90190-3} {\bibfield  {journal} {\bibinfo  {journal} {Journal of Physics and Chemistry of Solids}\ }\textbf {\bibinfo {volume} {26}},\ \bibinfo {pages} {1029} (\bibinfo {year} {1965})}\BibitemShut {NoStop}%
\bibitem [{\citenamefont {Ugeda}\ \emph {et~al.}(2016)\citenamefont {Ugeda}, \citenamefont {Bradley}, \citenamefont {Zhang}, \citenamefont {Onishi}, \citenamefont {Chen}, \citenamefont {Ruan}, \citenamefont {Ojeda-Aristizabal}, \citenamefont {Ryu}, \citenamefont {Edmonds}, \citenamefont {Tsai}, \citenamefont {Riss}, \citenamefont {Mo}, \citenamefont {Lee}, \citenamefont {Zettl}, \citenamefont {Hussain}, \citenamefont {Shen},\ and\ \citenamefont {Crommie}}]{Ugeda2016}%
  \BibitemOpen
  \bibfield  {author} {\bibinfo {author} {\bibfnamefont {M.~M.}\ \bibnamefont {Ugeda}}, \bibinfo {author} {\bibfnamefont {A.~J.}\ \bibnamefont {Bradley}}, \bibinfo {author} {\bibfnamefont {Y.}~\bibnamefont {Zhang}}, \bibinfo {author} {\bibfnamefont {S.}~\bibnamefont {Onishi}}, \bibinfo {author} {\bibfnamefont {Y.}~\bibnamefont {Chen}}, \bibinfo {author} {\bibfnamefont {W.}~\bibnamefont {Ruan}}, \bibinfo {author} {\bibfnamefont {C.}~\bibnamefont {Ojeda-Aristizabal}}, \bibinfo {author} {\bibfnamefont {H.}~\bibnamefont {Ryu}}, \bibinfo {author} {\bibfnamefont {M.~T.}\ \bibnamefont {Edmonds}}, \bibinfo {author} {\bibfnamefont {H.-Z.}\ \bibnamefont {Tsai}}, \bibinfo {author} {\bibfnamefont {A.}~\bibnamefont {Riss}}, \bibinfo {author} {\bibfnamefont {S.-K.}\ \bibnamefont {Mo}}, \bibinfo {author} {\bibfnamefont {D.}~\bibnamefont {Lee}}, \bibinfo {author} {\bibfnamefont {A.}~\bibnamefont {Zettl}}, \bibinfo {author} {\bibfnamefont {Z.}~\bibnamefont {Hussain}}, \bibinfo {author} {\bibfnamefont {Z.-X.}\ \bibnamefont
  {Shen}},\ and\ \bibinfo {author} {\bibfnamefont {M.~F.}\ \bibnamefont {Crommie}},\ }\bibfield  {title} {\bibinfo {title} {Characterization of collective ground states in single-layer {$\mathrm{NbSe}_{2}$}},\ }\href {https://doi.org/10.1038/nphys3527} {\bibfield  {journal} {\bibinfo  {journal} {Nature Physics}\ }\textbf {\bibinfo {volume} {12}},\ \bibinfo {pages} {92} (\bibinfo {year} {2016})}\BibitemShut {NoStop}%
\bibitem [{\citenamefont {Xi}\ \emph {et~al.}(2016)\citenamefont {Xi}, \citenamefont {Wang}, \citenamefont {Zhao}, \citenamefont {Park}, \citenamefont {Law}, \citenamefont {Berger}, \citenamefont {Forr{\'o}}, \citenamefont {Shan},\ and\ \citenamefont {Mak}}]{Xi2016}%
  \BibitemOpen
  \bibfield  {author} {\bibinfo {author} {\bibfnamefont {X.}~\bibnamefont {Xi}}, \bibinfo {author} {\bibfnamefont {Z.}~\bibnamefont {Wang}}, \bibinfo {author} {\bibfnamefont {W.}~\bibnamefont {Zhao}}, \bibinfo {author} {\bibfnamefont {J.-H.}\ \bibnamefont {Park}}, \bibinfo {author} {\bibfnamefont {K.~T.}\ \bibnamefont {Law}}, \bibinfo {author} {\bibfnamefont {H.}~\bibnamefont {Berger}}, \bibinfo {author} {\bibfnamefont {L.}~\bibnamefont {Forr{\'o}}}, \bibinfo {author} {\bibfnamefont {J.}~\bibnamefont {Shan}},\ and\ \bibinfo {author} {\bibfnamefont {K.~F.}\ \bibnamefont {Mak}},\ }\bibfield  {title} {\bibinfo {title} {Ising pairing in superconducting {$\mathrm{NbSe}_{2}$} atomic layers},\ }\href {https://doi.org/10.1038/nphys3538} {\bibfield  {journal} {\bibinfo  {journal} {Nature Physics}\ }\textbf {\bibinfo {volume} {12}},\ \bibinfo {pages} {139} (\bibinfo {year} {2016})}\BibitemShut {NoStop}%
\bibitem [{\citenamefont {Wang}\ \emph {et~al.}(2017)\citenamefont {Wang}, \citenamefont {Huang}, \citenamefont {Lin}, \citenamefont {Cui}, \citenamefont {Chen}, \citenamefont {Zhu}, \citenamefont {Liu}, \citenamefont {Zeng}, \citenamefont {Zhou}, \citenamefont {Yu}, \citenamefont {Wang}, \citenamefont {He}, \citenamefont {Tsang}, \citenamefont {Gao}, \citenamefont {Suenaga}, \citenamefont {Ma}, \citenamefont {Yang}, \citenamefont {Lu}, \citenamefont {Yu}, \citenamefont {Teo}, \citenamefont {Liu},\ and\ \citenamefont {Liu}}]{Wang2017}%
  \BibitemOpen
  \bibfield  {author} {\bibinfo {author} {\bibfnamefont {H.}~\bibnamefont {Wang}}, \bibinfo {author} {\bibfnamefont {X.}~\bibnamefont {Huang}}, \bibinfo {author} {\bibfnamefont {J.}~\bibnamefont {Lin}}, \bibinfo {author} {\bibfnamefont {J.}~\bibnamefont {Cui}}, \bibinfo {author} {\bibfnamefont {Y.}~\bibnamefont {Chen}}, \bibinfo {author} {\bibfnamefont {C.}~\bibnamefont {Zhu}}, \bibinfo {author} {\bibfnamefont {F.}~\bibnamefont {Liu}}, \bibinfo {author} {\bibfnamefont {Q.}~\bibnamefont {Zeng}}, \bibinfo {author} {\bibfnamefont {J.}~\bibnamefont {Zhou}}, \bibinfo {author} {\bibfnamefont {P.}~\bibnamefont {Yu}}, \bibinfo {author} {\bibfnamefont {X.}~\bibnamefont {Wang}}, \bibinfo {author} {\bibfnamefont {H.}~\bibnamefont {He}}, \bibinfo {author} {\bibfnamefont {S.~H.}\ \bibnamefont {Tsang}}, \bibinfo {author} {\bibfnamefont {W.}~\bibnamefont {Gao}}, \bibinfo {author} {\bibfnamefont {K.}~\bibnamefont {Suenaga}}, \bibinfo {author} {\bibfnamefont {F.}~\bibnamefont {Ma}}, \bibinfo {author} {\bibfnamefont
  {C.}~\bibnamefont {Yang}}, \bibinfo {author} {\bibfnamefont {L.}~\bibnamefont {Lu}}, \bibinfo {author} {\bibfnamefont {T.}~\bibnamefont {Yu}}, \bibinfo {author} {\bibfnamefont {E.~H.~T.}\ \bibnamefont {Teo}}, \bibinfo {author} {\bibfnamefont {G.}~\bibnamefont {Liu}},\ and\ \bibinfo {author} {\bibfnamefont {Z.}~\bibnamefont {Liu}},\ }\bibfield  {title} {\bibinfo {title} {High-quality monolayer superconductor {$\mathrm{NbSe}_{2}$} grown by chemical vapour deposition},\ }\href {https://doi.org/10.1038/s41467-017-00427-5} {\bibfield  {journal} {\bibinfo  {journal} {Nature Communications}\ }\textbf {\bibinfo {volume} {8}},\ \bibinfo {pages} {394} (\bibinfo {year} {2017})}\BibitemShut {NoStop}%
\bibitem [{\citenamefont {Xing}\ \emph {et~al.}(2017)\citenamefont {Xing}, \citenamefont {Zhao}, \citenamefont {Shan}, \citenamefont {Zheng}, \citenamefont {Zhang}, \citenamefont {Fu}, \citenamefont {Liu}, \citenamefont {Tian}, \citenamefont {Xi}, \citenamefont {Liu}, \citenamefont {Feng}, \citenamefont {Lin}, \citenamefont {Ji}, \citenamefont {Chen}, \citenamefont {Xue},\ and\ \citenamefont {Wang}}]{Xing2017}%
  \BibitemOpen
  \bibfield  {author} {\bibinfo {author} {\bibfnamefont {Y.}~\bibnamefont {Xing}}, \bibinfo {author} {\bibfnamefont {K.}~\bibnamefont {Zhao}}, \bibinfo {author} {\bibfnamefont {P.}~\bibnamefont {Shan}}, \bibinfo {author} {\bibfnamefont {F.}~\bibnamefont {Zheng}}, \bibinfo {author} {\bibfnamefont {Y.}~\bibnamefont {Zhang}}, \bibinfo {author} {\bibfnamefont {H.}~\bibnamefont {Fu}}, \bibinfo {author} {\bibfnamefont {Y.}~\bibnamefont {Liu}}, \bibinfo {author} {\bibfnamefont {M.}~\bibnamefont {Tian}}, \bibinfo {author} {\bibfnamefont {C.}~\bibnamefont {Xi}}, \bibinfo {author} {\bibfnamefont {H.}~\bibnamefont {Liu}}, \bibinfo {author} {\bibfnamefont {J.}~\bibnamefont {Feng}}, \bibinfo {author} {\bibfnamefont {X.}~\bibnamefont {Lin}}, \bibinfo {author} {\bibfnamefont {S.}~\bibnamefont {Ji}}, \bibinfo {author} {\bibfnamefont {X.}~\bibnamefont {Chen}}, \bibinfo {author} {\bibfnamefont {Q.-K.}\ \bibnamefont {Xue}},\ and\ \bibinfo {author} {\bibfnamefont {J.}~\bibnamefont {Wang}},\ }\bibfield  {title} {\bibinfo {title}
  {Ising superconductivity and quantum phase transition in macro-size monolayer {$\mathrm{NbSe}_{2}$}},\ }\href {https://doi.org/10.1021/acs.nanolett.7b03026} {\bibfield  {journal} {\bibinfo  {journal} {Nano Letters}\ }\textbf {\bibinfo {volume} {17}},\ \bibinfo {pages} {6802} (\bibinfo {year} {2017})}\BibitemShut {NoStop}%
\bibitem [{\citenamefont {Zhao}\ \emph {et~al.}(2019)\citenamefont {Zhao}, \citenamefont {Lin}, \citenamefont {Xiao}, \citenamefont {Huang}, \citenamefont {Yao}, \citenamefont {Yan}, \citenamefont {Xing}, \citenamefont {Zhang}, \citenamefont {Li}, \citenamefont {Hoshino}, \citenamefont {Wang}, \citenamefont {Zhou}, \citenamefont {Gu}, \citenamefont {Bahramy}, \citenamefont {Yao}, \citenamefont {Nagaosa}, \citenamefont {Xue}, \citenamefont {Law}, \citenamefont {Chen},\ and\ \citenamefont {Ji}}]{Zhao2019}%
  \BibitemOpen
  \bibfield  {author} {\bibinfo {author} {\bibfnamefont {K.}~\bibnamefont {Zhao}}, \bibinfo {author} {\bibfnamefont {H.}~\bibnamefont {Lin}}, \bibinfo {author} {\bibfnamefont {X.}~\bibnamefont {Xiao}}, \bibinfo {author} {\bibfnamefont {W.}~\bibnamefont {Huang}}, \bibinfo {author} {\bibfnamefont {W.}~\bibnamefont {Yao}}, \bibinfo {author} {\bibfnamefont {M.}~\bibnamefont {Yan}}, \bibinfo {author} {\bibfnamefont {Y.}~\bibnamefont {Xing}}, \bibinfo {author} {\bibfnamefont {Q.}~\bibnamefont {Zhang}}, \bibinfo {author} {\bibfnamefont {Z.-X.}\ \bibnamefont {Li}}, \bibinfo {author} {\bibfnamefont {S.}~\bibnamefont {Hoshino}}, \bibinfo {author} {\bibfnamefont {J.}~\bibnamefont {Wang}}, \bibinfo {author} {\bibfnamefont {S.}~\bibnamefont {Zhou}}, \bibinfo {author} {\bibfnamefont {L.}~\bibnamefont {Gu}}, \bibinfo {author} {\bibfnamefont {M.~S.}\ \bibnamefont {Bahramy}}, \bibinfo {author} {\bibfnamefont {H.}~\bibnamefont {Yao}}, \bibinfo {author} {\bibfnamefont {N.}~\bibnamefont {Nagaosa}}, \bibinfo {author}
  {\bibfnamefont {Q.-K.}\ \bibnamefont {Xue}}, \bibinfo {author} {\bibfnamefont {K.~T.}\ \bibnamefont {Law}}, \bibinfo {author} {\bibfnamefont {X.}~\bibnamefont {Chen}},\ and\ \bibinfo {author} {\bibfnamefont {S.-H.}\ \bibnamefont {Ji}},\ }\bibfield  {title} {\bibinfo {title} {Disorder-induced multifractal superconductivity in monolayer niobium dichalcogenides},\ }\href {https://doi.org/10.1038/s41567-019-0570-0} {\bibfield  {journal} {\bibinfo  {journal} {Nature Physics}\ }\textbf {\bibinfo {volume} {15}},\ \bibinfo {pages} {904} (\bibinfo {year} {2019})}\BibitemShut {NoStop}%
\bibitem [{\citenamefont {Chen}\ \emph {et~al.}(2020)\citenamefont {Chen}, \citenamefont {Wu}, \citenamefont {Xu}, \citenamefont {Cong}, \citenamefont {Li}, \citenamefont {Feng}, \citenamefont {Zhang}, \citenamefont {Zou}, \citenamefont {Shang}, \citenamefont {Yang}, \citenamefont {Loh}, \citenamefont {Huang},\ and\ \citenamefont {Yu}}]{Chen2020}%
  \BibitemOpen
  \bibfield  {author} {\bibinfo {author} {\bibfnamefont {Y.}~\bibnamefont {Chen}}, \bibinfo {author} {\bibfnamefont {L.}~\bibnamefont {Wu}}, \bibinfo {author} {\bibfnamefont {H.}~\bibnamefont {Xu}}, \bibinfo {author} {\bibfnamefont {C.}~\bibnamefont {Cong}}, \bibinfo {author} {\bibfnamefont {S.}~\bibnamefont {Li}}, \bibinfo {author} {\bibfnamefont {S.}~\bibnamefont {Feng}}, \bibinfo {author} {\bibfnamefont {H.}~\bibnamefont {Zhang}}, \bibinfo {author} {\bibfnamefont {C.}~\bibnamefont {Zou}}, \bibinfo {author} {\bibfnamefont {J.}~\bibnamefont {Shang}}, \bibinfo {author} {\bibfnamefont {S.~A.}\ \bibnamefont {Yang}}, \bibinfo {author} {\bibfnamefont {K.~P.}\ \bibnamefont {Loh}}, \bibinfo {author} {\bibfnamefont {W.}~\bibnamefont {Huang}},\ and\ \bibinfo {author} {\bibfnamefont {T.}~\bibnamefont {Yu}},\ }\bibfield  {title} {\bibinfo {title} {Visualizing the anomalous charge density wave states in graphene/{$\mathrm{NbSe}_{2}$} heterostructures},\ }\href {https://doi.org/https://doi.org/10.1002/adma.202003746}
  {\bibfield  {journal} {\bibinfo  {journal} {Advanced Materials}\ }\textbf {\bibinfo {volume} {32}},\ \bibinfo {pages} {2003746} (\bibinfo {year} {2020})}\BibitemShut {NoStop}%
\bibitem [{\citenamefont {Dreher}\ \emph {et~al.}(2021)\citenamefont {Dreher}, \citenamefont {Wan}, \citenamefont {Chikina}, \citenamefont {Bianchi}, \citenamefont {Guo}, \citenamefont {Harsh}, \citenamefont {Mañas-Valero}, \citenamefont {Coronado}, \citenamefont {Martínez-Galera}, \citenamefont {Hofmann}, \citenamefont {Miwa},\ and\ \citenamefont {Ugeda}}]{Dreher2021}%
  \BibitemOpen
  \bibfield  {author} {\bibinfo {author} {\bibfnamefont {P.}~\bibnamefont {Dreher}}, \bibinfo {author} {\bibfnamefont {W.}~\bibnamefont {Wan}}, \bibinfo {author} {\bibfnamefont {A.}~\bibnamefont {Chikina}}, \bibinfo {author} {\bibfnamefont {M.}~\bibnamefont {Bianchi}}, \bibinfo {author} {\bibfnamefont {H.}~\bibnamefont {Guo}}, \bibinfo {author} {\bibfnamefont {R.}~\bibnamefont {Harsh}}, \bibinfo {author} {\bibfnamefont {S.}~\bibnamefont {Mañas-Valero}}, \bibinfo {author} {\bibfnamefont {E.}~\bibnamefont {Coronado}}, \bibinfo {author} {\bibfnamefont {A.~J.}\ \bibnamefont {Martínez-Galera}}, \bibinfo {author} {\bibfnamefont {P.}~\bibnamefont {Hofmann}}, \bibinfo {author} {\bibfnamefont {J.~A.}\ \bibnamefont {Miwa}},\ and\ \bibinfo {author} {\bibfnamefont {M.~M.}\ \bibnamefont {Ugeda}},\ }\bibfield  {title} {\bibinfo {title} {Proximity effects on the charge density wave order and superconductivity in single-layer {$\mathrm{NbSe}_{2}$}},\ }\href {https://doi.org/10.1021/acsnano.1c06012} {\bibfield  {journal}
  {\bibinfo  {journal} {ACS Nano}\ }\textbf {\bibinfo {volume} {15}},\ \bibinfo {pages} {19430} (\bibinfo {year} {2021})}\BibitemShut {NoStop}%
\bibitem [{\citenamefont {Ganguli}\ \emph {et~al.}(2022)\citenamefont {Ganguli}, \citenamefont {Vaňo}, \citenamefont {Kezilebieke}, \citenamefont {Lado},\ and\ \citenamefont {Liljeroth}}]{Ganguli2022}%
  \BibitemOpen
  \bibfield  {author} {\bibinfo {author} {\bibfnamefont {S.~C.}\ \bibnamefont {Ganguli}}, \bibinfo {author} {\bibfnamefont {V.}~\bibnamefont {Vaňo}}, \bibinfo {author} {\bibfnamefont {S.}~\bibnamefont {Kezilebieke}}, \bibinfo {author} {\bibfnamefont {J.~L.}\ \bibnamefont {Lado}},\ and\ \bibinfo {author} {\bibfnamefont {P.}~\bibnamefont {Liljeroth}},\ }\bibfield  {title} {\bibinfo {title} {Confinement-engineered superconductor to correlated-insulator transition in a van der waals monolayer},\ }\href {https://doi.org/10.1021/acs.nanolett.1c03491} {\bibfield  {journal} {\bibinfo  {journal} {Nano Letters}\ }\textbf {\bibinfo {volume} {22}},\ \bibinfo {pages} {1845} (\bibinfo {year} {2022})}\BibitemShut {NoStop}%
\bibitem [{\citenamefont {Machida}\ \emph {et~al.}(2018)\citenamefont {Machida}, \citenamefont {Kohsaka},\ and\ \citenamefont {Hanaguri}}]{Machida2018}%
  \BibitemOpen
  \bibfield  {author} {\bibinfo {author} {\bibfnamefont {T.}~\bibnamefont {Machida}}, \bibinfo {author} {\bibfnamefont {Y.}~\bibnamefont {Kohsaka}},\ and\ \bibinfo {author} {\bibfnamefont {T.}~\bibnamefont {Hanaguri}},\ }\bibfield  {title} {\bibinfo {title} {{A scanning tunneling microscope for spectroscopic imaging below 90 mK in magnetic fields up to 17.5 {T}}},\ }\href {https://doi.org/10.1063/1.5049619} {\bibfield  {journal} {\bibinfo  {journal} {Review of Scientific Instruments}\ }\textbf {\bibinfo {volume} {89}},\ \bibinfo {pages} {093707} (\bibinfo {year} {2018})}\BibitemShut {NoStop}%
\bibitem [{Sup()}]{Supplement}%
  \BibitemOpen
  \href@noop {} {}\bibinfo {note} {Materials and methods are available as supplementary materials.}\BibitemShut {Stop}%
\bibitem [{\citenamefont {Goler}\ \emph {et~al.}(2013)\citenamefont {Goler}, \citenamefont {Coletti}, \citenamefont {Piazza}, \citenamefont {Pingue}, \citenamefont {Colangelo}, \citenamefont {Pellegrini}, \citenamefont {Emtsev}, \citenamefont {Forti}, \citenamefont {Starke}, \citenamefont {Beltram},\ and\ \citenamefont {Heun}}]{Goler2013}%
  \BibitemOpen
  \bibfield  {author} {\bibinfo {author} {\bibfnamefont {S.}~\bibnamefont {Goler}}, \bibinfo {author} {\bibfnamefont {C.}~\bibnamefont {Coletti}}, \bibinfo {author} {\bibfnamefont {V.}~\bibnamefont {Piazza}}, \bibinfo {author} {\bibfnamefont {P.}~\bibnamefont {Pingue}}, \bibinfo {author} {\bibfnamefont {F.}~\bibnamefont {Colangelo}}, \bibinfo {author} {\bibfnamefont {V.}~\bibnamefont {Pellegrini}}, \bibinfo {author} {\bibfnamefont {K.~V.}\ \bibnamefont {Emtsev}}, \bibinfo {author} {\bibfnamefont {S.}~\bibnamefont {Forti}}, \bibinfo {author} {\bibfnamefont {U.}~\bibnamefont {Starke}}, \bibinfo {author} {\bibfnamefont {F.}~\bibnamefont {Beltram}},\ and\ \bibinfo {author} {\bibfnamefont {S.}~\bibnamefont {Heun}},\ }\bibfield  {title} {\bibinfo {title} {Revealing the atomic structure of the buffer layer between {SiC(0001)} and epitaxial graphene},\ }\href {https://doi.org/https://doi.org/10.1016/j.carbon.2012.08.050} {\bibfield  {journal} {\bibinfo  {journal} {Carbon}\ }\textbf {\bibinfo {volume} {51}},\ \bibinfo
  {pages} {249} (\bibinfo {year} {2013})}\BibitemShut {NoStop}%
\bibitem [{\citenamefont {Riedl}\ \emph {et~al.}(2010)\citenamefont {Riedl}, \citenamefont {Coletti},\ and\ \citenamefont {Starke}}]{Riedl2010}%
  \BibitemOpen
  \bibfield  {author} {\bibinfo {author} {\bibfnamefont {C.}~\bibnamefont {Riedl}}, \bibinfo {author} {\bibfnamefont {C.}~\bibnamefont {Coletti}},\ and\ \bibinfo {author} {\bibfnamefont {U.}~\bibnamefont {Starke}},\ }\bibfield  {title} {\bibinfo {title} {Structural and electronic properties of epitaxial graphene on {SiC(0001)}: a review of growth, characterization, transfer doping and hydrogen intercalation},\ }\href {https://doi.org/10.1088/0022-3727/43/37/374009} {\bibfield  {journal} {\bibinfo  {journal} {Journal of Physics D: Applied Physics}\ }\textbf {\bibinfo {volume} {43}},\ \bibinfo {pages} {374009} (\bibinfo {year} {2010})}\BibitemShut {NoStop}%
\bibitem [{\citenamefont {Kohsaka}\ \emph {et~al.}(2007)\citenamefont {Kohsaka}, \citenamefont {Taylor}, \citenamefont {Fujita}, \citenamefont {Schmidt}, \citenamefont {Lupien}, \citenamefont {Hanaguri}, \citenamefont {Azuma}, \citenamefont {Takano}, \citenamefont {Eisaki}, \citenamefont {Takagi}, \citenamefont {Uchida},\ and\ \citenamefont {Davis}}]{Kohsaka2007}%
  \BibitemOpen
  \bibfield  {author} {\bibinfo {author} {\bibfnamefont {Y.}~\bibnamefont {Kohsaka}}, \bibinfo {author} {\bibfnamefont {C.}~\bibnamefont {Taylor}}, \bibinfo {author} {\bibfnamefont {K.}~\bibnamefont {Fujita}}, \bibinfo {author} {\bibfnamefont {A.}~\bibnamefont {Schmidt}}, \bibinfo {author} {\bibfnamefont {C.}~\bibnamefont {Lupien}}, \bibinfo {author} {\bibfnamefont {T.}~\bibnamefont {Hanaguri}}, \bibinfo {author} {\bibfnamefont {M.}~\bibnamefont {Azuma}}, \bibinfo {author} {\bibfnamefont {M.}~\bibnamefont {Takano}}, \bibinfo {author} {\bibfnamefont {H.}~\bibnamefont {Eisaki}}, \bibinfo {author} {\bibfnamefont {H.}~\bibnamefont {Takagi}}, \bibinfo {author} {\bibfnamefont {S.}~\bibnamefont {Uchida}},\ and\ \bibinfo {author} {\bibfnamefont {J.~C.}\ \bibnamefont {Davis}},\ }\bibfield  {title} {\bibinfo {title} {An intrinsic bond-centered electronic glass with unidirectional domains in underdoped cuprates},\ }\href {https://doi.org/10.1126/science.1138584} {\bibfield  {journal} {\bibinfo  {journal} {Science}\
  }\textbf {\bibinfo {volume} {315}},\ \bibinfo {pages} {1380} (\bibinfo {year} {2007})}\BibitemShut {NoStop}%
\bibitem [{\citenamefont {Pham}\ \emph {et~al.}(2022)\citenamefont {Pham}, \citenamefont {Vancs{\'o}}, \citenamefont {Szendr{\H{o}}}, \citenamefont {Palot{\'a}s}, \citenamefont {Castelino}, \citenamefont {Bouatou}, \citenamefont {Chacon}, \citenamefont {Henrard}, \citenamefont {Lagoute},\ and\ \citenamefont {Sporken}}]{Pham2022}%
  \BibitemOpen
  \bibfield  {author} {\bibinfo {author} {\bibfnamefont {T.~T.}\ \bibnamefont {Pham}}, \bibinfo {author} {\bibfnamefont {P.}~\bibnamefont {Vancs{\'o}}}, \bibinfo {author} {\bibfnamefont {M.}~\bibnamefont {Szendr{\H{o}}}}, \bibinfo {author} {\bibfnamefont {K.}~\bibnamefont {Palot{\'a}s}}, \bibinfo {author} {\bibfnamefont {R.}~\bibnamefont {Castelino}}, \bibinfo {author} {\bibfnamefont {M.}~\bibnamefont {Bouatou}}, \bibinfo {author} {\bibfnamefont {C.}~\bibnamefont {Chacon}}, \bibinfo {author} {\bibfnamefont {L.}~\bibnamefont {Henrard}}, \bibinfo {author} {\bibfnamefont {J.}~\bibnamefont {Lagoute}},\ and\ \bibinfo {author} {\bibfnamefont {R.}~\bibnamefont {Sporken}},\ }\bibfield  {title} {\bibinfo {title} {Higher-indexed moir{\'e} patterns and surface states of {$\mathrm{MoTe}_{2}$}/graphene heterostructure grown by molecular beam epitaxy},\ }\href {https://doi.org/10.1038/s41699-022-00321-9} {\bibfield  {journal} {\bibinfo  {journal} {npj 2D Materials and Applications}\ }\textbf {\bibinfo {volume} {6}},\
  \bibinfo {pages} {48} (\bibinfo {year} {2022})}\BibitemShut {NoStop}%
\bibitem [{\citenamefont {Liu}\ \emph {et~al.}(2013)\citenamefont {Liu}, \citenamefont {Shan}, \citenamefont {Yao}, \citenamefont {Yao},\ and\ \citenamefont {Xiao}}]{Liu2013}%
  \BibitemOpen
  \bibfield  {author} {\bibinfo {author} {\bibfnamefont {G.-B.}\ \bibnamefont {Liu}}, \bibinfo {author} {\bibfnamefont {W.-Y.}\ \bibnamefont {Shan}}, \bibinfo {author} {\bibfnamefont {Y.}~\bibnamefont {Yao}}, \bibinfo {author} {\bibfnamefont {W.}~\bibnamefont {Yao}},\ and\ \bibinfo {author} {\bibfnamefont {D.}~\bibnamefont {Xiao}},\ }\bibfield  {title} {\bibinfo {title} {Three-band tight-binding model for monolayers of {group-VIB} transition metal dichalcogenides},\ }\href {https://doi.org/10.1103/PhysRevB.88.085433} {\bibfield  {journal} {\bibinfo  {journal} {Phys. Rev. B}\ }\textbf {\bibinfo {volume} {88}},\ \bibinfo {pages} {085433} (\bibinfo {year} {2013})}\BibitemShut {NoStop}%
\bibitem [{\citenamefont {Arguello}\ \emph {et~al.}(2015)\citenamefont {Arguello}, \citenamefont {Rosenthal}, \citenamefont {Andrade}, \citenamefont {Jin}, \citenamefont {Yeh}, \citenamefont {Zaki}, \citenamefont {Jia}, \citenamefont {Cava}, \citenamefont {Fernandes}, \citenamefont {Millis}, \citenamefont {Valla}, \citenamefont {Osgood},\ and\ \citenamefont {Pasupathy}}]{Arguello2015}%
  \BibitemOpen
  \bibfield  {author} {\bibinfo {author} {\bibfnamefont {C.~J.}\ \bibnamefont {Arguello}}, \bibinfo {author} {\bibfnamefont {E.~P.}\ \bibnamefont {Rosenthal}}, \bibinfo {author} {\bibfnamefont {E.~F.}\ \bibnamefont {Andrade}}, \bibinfo {author} {\bibfnamefont {W.}~\bibnamefont {Jin}}, \bibinfo {author} {\bibfnamefont {P.~C.}\ \bibnamefont {Yeh}}, \bibinfo {author} {\bibfnamefont {N.}~\bibnamefont {Zaki}}, \bibinfo {author} {\bibfnamefont {S.}~\bibnamefont {Jia}}, \bibinfo {author} {\bibfnamefont {R.~J.}\ \bibnamefont {Cava}}, \bibinfo {author} {\bibfnamefont {R.~M.}\ \bibnamefont {Fernandes}}, \bibinfo {author} {\bibfnamefont {A.~J.}\ \bibnamefont {Millis}}, \bibinfo {author} {\bibfnamefont {T.}~\bibnamefont {Valla}}, \bibinfo {author} {\bibfnamefont {R.~M.}\ \bibnamefont {Osgood}},\ and\ \bibinfo {author} {\bibfnamefont {A.~N.}\ \bibnamefont {Pasupathy}},\ }\bibfield  {title} {\bibinfo {title} {Quasiparticle interference, quasiparticle interactions, and the origin of the charge density wave in
  {$2H\text{\ensuremath{-}}{\mathrm{NbSe}}_{2}$}},\ }\href {https://doi.org/10.1103/PhysRevLett.114.037001} {\bibfield  {journal} {\bibinfo  {journal} {Phys. Rev. Lett.}\ }\textbf {\bibinfo {volume} {114}},\ \bibinfo {pages} {037001} (\bibinfo {year} {2015})}\BibitemShut {NoStop}%
\bibitem [{\citenamefont {Gao}\ \emph {et~al.}(2018)\citenamefont {Gao}, \citenamefont {Flicker}, \citenamefont {Sankar}, \citenamefont {Zhao}, \citenamefont {Ren}, \citenamefont {Rachmilowitz}, \citenamefont {Balachandar}, \citenamefont {Chou}, \citenamefont {Burch}, \citenamefont {Wang}, \citenamefont {van Wezel},\ and\ \citenamefont {Zeljkovic}}]{Gao2018}%
  \BibitemOpen
  \bibfield  {author} {\bibinfo {author} {\bibfnamefont {S.}~\bibnamefont {Gao}}, \bibinfo {author} {\bibfnamefont {F.}~\bibnamefont {Flicker}}, \bibinfo {author} {\bibfnamefont {R.}~\bibnamefont {Sankar}}, \bibinfo {author} {\bibfnamefont {H.}~\bibnamefont {Zhao}}, \bibinfo {author} {\bibfnamefont {Z.}~\bibnamefont {Ren}}, \bibinfo {author} {\bibfnamefont {B.}~\bibnamefont {Rachmilowitz}}, \bibinfo {author} {\bibfnamefont {S.}~\bibnamefont {Balachandar}}, \bibinfo {author} {\bibfnamefont {F.}~\bibnamefont {Chou}}, \bibinfo {author} {\bibfnamefont {K.~S.}\ \bibnamefont {Burch}}, \bibinfo {author} {\bibfnamefont {Z.}~\bibnamefont {Wang}}, \bibinfo {author} {\bibfnamefont {J.}~\bibnamefont {van Wezel}},\ and\ \bibinfo {author} {\bibfnamefont {I.}~\bibnamefont {Zeljkovic}},\ }\bibfield  {title} {\bibinfo {title} {Atomic-scale strain manipulation of a charge density wave},\ }\href {https://doi.org/10.1073/pnas.1718931115} {\bibfield  {journal} {\bibinfo  {journal} {Proceedings of the National Academy of
  Sciences}\ }\textbf {\bibinfo {volume} {115}},\ \bibinfo {pages} {6986} (\bibinfo {year} {2018})}\BibitemShut {NoStop}%
\bibitem [{\citenamefont {Brun}\ \emph {et~al.}(2012)\citenamefont {Brun}, \citenamefont {M\"uller}, \citenamefont {Hong}, \citenamefont {Patthey}, \citenamefont {Flindt},\ and\ \citenamefont {Schneider}}]{Brun2012}%
  \BibitemOpen
  \bibfield  {author} {\bibinfo {author} {\bibfnamefont {C.}~\bibnamefont {Brun}}, \bibinfo {author} {\bibfnamefont {K.~H.}\ \bibnamefont {M\"uller}}, \bibinfo {author} {\bibfnamefont {I.-P.}\ \bibnamefont {Hong}}, \bibinfo {author} {\bibfnamefont {F.~m.~c.}\ \bibnamefont {Patthey}}, \bibinfo {author} {\bibfnamefont {C.}~\bibnamefont {Flindt}},\ and\ \bibinfo {author} {\bibfnamefont {W.-D.}\ \bibnamefont {Schneider}},\ }\bibfield  {title} {\bibinfo {title} {Dynamical coulomb blockade observed in nanosized electrical contacts},\ }\href {https://doi.org/10.1103/PhysRevLett.108.126802} {\bibfield  {journal} {\bibinfo  {journal} {Phys. Rev. Lett.}\ }\textbf {\bibinfo {volume} {108}},\ \bibinfo {pages} {126802} (\bibinfo {year} {2012})}\BibitemShut {NoStop}%
\bibitem [{\citenamefont {Devoret}\ \emph {et~al.}(1990)\citenamefont {Devoret}, \citenamefont {Esteve}, \citenamefont {Grabert}, \citenamefont {Ingold}, \citenamefont {Pothier},\ and\ \citenamefont {Urbina}}]{Devoret1990}%
  \BibitemOpen
  \bibfield  {author} {\bibinfo {author} {\bibfnamefont {M.~H.}\ \bibnamefont {Devoret}}, \bibinfo {author} {\bibfnamefont {D.}~\bibnamefont {Esteve}}, \bibinfo {author} {\bibfnamefont {H.}~\bibnamefont {Grabert}}, \bibinfo {author} {\bibfnamefont {G.-L.}\ \bibnamefont {Ingold}}, \bibinfo {author} {\bibfnamefont {H.}~\bibnamefont {Pothier}},\ and\ \bibinfo {author} {\bibfnamefont {C.}~\bibnamefont {Urbina}},\ }\bibfield  {title} {\bibinfo {title} {Effect of the electromagnetic environment on the coulomb blockade in ultrasmall tunnel junctions},\ }\href {https://doi.org/10.1103/PhysRevLett.64.1824} {\bibfield  {journal} {\bibinfo  {journal} {Phys. Rev. Lett.}\ }\textbf {\bibinfo {volume} {64}},\ \bibinfo {pages} {1824} (\bibinfo {year} {1990})}\BibitemShut {NoStop}%
\bibitem [{\citenamefont {Ingold}\ and\ \citenamefont {Nazarov}(1992)}]{Ingold1992}%
  \BibitemOpen
  \bibfield  {author} {\bibinfo {author} {\bibfnamefont {G.-L.}\ \bibnamefont {Ingold}}\ and\ \bibinfo {author} {\bibfnamefont {Y.~V.}\ \bibnamefont {Nazarov}},\ }\bibinfo {title} {Charge tunneling rates in ultrasmall junctions},\ in\ \href {https://doi.org/10.1007/978-1-4757-2166-9_2} {\emph {\bibinfo {booktitle} {Single Charge Tunneling: Coulomb Blockade Phenomena In Nanostructures}}}\ (\bibinfo  {publisher} {Springer US},\ \bibinfo {address} {Boston, MA},\ \bibinfo {year} {1992})\ pp.\ \bibinfo {pages} {21--107}\BibitemShut {NoStop}%
\bibitem [{\citenamefont {Ast}\ \emph {et~al.}(2016)\citenamefont {Ast}, \citenamefont {J{\"a}ck}, \citenamefont {Senkpiel}, \citenamefont {Eltschka}, \citenamefont {Etzkorn}, \citenamefont {Ankerhold},\ and\ \citenamefont {Kern}}]{Ast2016}%
  \BibitemOpen
  \bibfield  {author} {\bibinfo {author} {\bibfnamefont {C.~R.}\ \bibnamefont {Ast}}, \bibinfo {author} {\bibfnamefont {B.}~\bibnamefont {J{\"a}ck}}, \bibinfo {author} {\bibfnamefont {J.}~\bibnamefont {Senkpiel}}, \bibinfo {author} {\bibfnamefont {M.}~\bibnamefont {Eltschka}}, \bibinfo {author} {\bibfnamefont {M.}~\bibnamefont {Etzkorn}}, \bibinfo {author} {\bibfnamefont {J.}~\bibnamefont {Ankerhold}},\ and\ \bibinfo {author} {\bibfnamefont {K.}~\bibnamefont {Kern}},\ }\bibfield  {title} {\bibinfo {title} {Sensing the quantum limit in scanning tunnelling spectroscopy},\ }\href {https://doi.org/10.1038/ncomms13009} {\bibfield  {journal} {\bibinfo  {journal} {Nature Communications}\ }\textbf {\bibinfo {volume} {7}},\ \bibinfo {pages} {13009} (\bibinfo {year} {2016})}\BibitemShut {NoStop}%
\bibitem [{\citenamefont {Gor'kov}\ and\ \citenamefont {Rashba}(2001)}]{Gorkov2001}%
  \BibitemOpen
  \bibfield  {author} {\bibinfo {author} {\bibfnamefont {L.~P.}\ \bibnamefont {Gor'kov}}\ and\ \bibinfo {author} {\bibfnamefont {E.~I.}\ \bibnamefont {Rashba}},\ }\bibfield  {title} {\bibinfo {title} {Superconducting {2D} system with lifted spin degeneracy: Mixed singlet-triplet state},\ }\href {https://doi.org/10.1103/PhysRevLett.87.037004} {\bibfield  {journal} {\bibinfo  {journal} {Phys. Rev. Lett.}\ }\textbf {\bibinfo {volume} {87}},\ \bibinfo {pages} {037004} (\bibinfo {year} {2001})}\BibitemShut {NoStop}%
\bibitem [{\citenamefont {Gani}\ \emph {et~al.}(2019)\citenamefont {Gani}, \citenamefont {Steinberg},\ and\ \citenamefont {Rossi}}]{Gani2019}%
  \BibitemOpen
  \bibfield  {author} {\bibinfo {author} {\bibfnamefont {Y.~S.}\ \bibnamefont {Gani}}, \bibinfo {author} {\bibfnamefont {H.}~\bibnamefont {Steinberg}},\ and\ \bibinfo {author} {\bibfnamefont {E.}~\bibnamefont {Rossi}},\ }\bibfield  {title} {\bibinfo {title} {Superconductivity in twisted graphene {${\mathrm{NbSe}}_{2}$} heterostructures},\ }\href {https://doi.org/10.1103/PhysRevB.99.235404} {\bibfield  {journal} {\bibinfo  {journal} {Phys. Rev. B}\ }\textbf {\bibinfo {volume} {99}},\ \bibinfo {pages} {235404} (\bibinfo {year} {2019})}\BibitemShut {NoStop}%
\bibitem [{\citenamefont {Li}\ and\ \citenamefont {Koshino}(2019)}]{Li2019}%
  \BibitemOpen
  \bibfield  {author} {\bibinfo {author} {\bibfnamefont {Y.}~\bibnamefont {Li}}\ and\ \bibinfo {author} {\bibfnamefont {M.}~\bibnamefont {Koshino}},\ }\bibfield  {title} {\bibinfo {title} {Twist-angle dependence of the proximity spin-orbit coupling in graphene on transition-metal dichalcogenides},\ }\href {https://doi.org/10.1103/PhysRevB.99.075438} {\bibfield  {journal} {\bibinfo  {journal} {Phys. Rev. B}\ }\textbf {\bibinfo {volume} {99}},\ \bibinfo {pages} {075438} (\bibinfo {year} {2019})}\BibitemShut {NoStop}%
\bibitem [{\citenamefont {Coletti}\ \emph {et~al.}(2010)\citenamefont {Coletti}, \citenamefont {Riedl}, \citenamefont {Lee}, \citenamefont {Krauss}, \citenamefont {Patthey}, \citenamefont {von Klitzing}, \citenamefont {Smet},\ and\ \citenamefont {Starke}}]{Coletti2010}%
  \BibitemOpen
  \bibfield  {author} {\bibinfo {author} {\bibfnamefont {C.}~\bibnamefont {Coletti}}, \bibinfo {author} {\bibfnamefont {C.}~\bibnamefont {Riedl}}, \bibinfo {author} {\bibfnamefont {D.~S.}\ \bibnamefont {Lee}}, \bibinfo {author} {\bibfnamefont {B.}~\bibnamefont {Krauss}}, \bibinfo {author} {\bibfnamefont {L.}~\bibnamefont {Patthey}}, \bibinfo {author} {\bibfnamefont {K.}~\bibnamefont {von Klitzing}}, \bibinfo {author} {\bibfnamefont {J.~H.}\ \bibnamefont {Smet}},\ and\ \bibinfo {author} {\bibfnamefont {U.}~\bibnamefont {Starke}},\ }\bibfield  {title} {\bibinfo {title} {Charge neutrality and band-gap tuning of epitaxial graphene on {SiC} by molecular doping},\ }\href {https://doi.org/10.1103/PhysRevB.81.235401} {\bibfield  {journal} {\bibinfo  {journal} {Phys. Rev. B}\ }\textbf {\bibinfo {volume} {81}},\ \bibinfo {pages} {235401} (\bibinfo {year} {2010})}\BibitemShut {NoStop}%
\end{thebibliography}%

\section*{Acknowledgments}

The authors thank C. J. Butler, M. Nakano, K. Sugawara, Y. Okada, and Y. Hasegawa for valuable discussions and comments.

\textbf{Funding:}
M.N. acknowledges support from RIKEN's SPDR fellowship.
This work was supported by JSPS KAKENHI Grant Numbers JP19H05824, JP21K18145, JP22K18696, JP22K20362,  JP22H01181, JP22H04933, JP23H01134, JP23K13067, JP23K17353 and PRESTO project from Japan Science and Technology Agency (No. JPMJPR19L8).
 
\textbf{Author contributions:}
M.N. prepared samples, carried out SI-STM measurements, and analyzed the data with assistance from T.M. S.A. and Y.Y. contributed to construct the sextet model.
T.H. supervised the project.
M.N. and T.H. wrote the manuscript.
All authors discussed the results and contributed to finalize the manuscript.

\textbf{Competing interests:}
The authors declare that they have no competing interest.

\textbf{Data Availability:}
All data needed to evaluate the conclusions in this paper are present in the paper and/or the Supplementary Materials.
Additional data related to this paper may be requested from the authors.

\clearpage
\newgeometry{top=1.5cm, left=1.0cm, right=1.0cm}
\begin{figure*}
\centering
\includegraphics[width=0.85\linewidth]{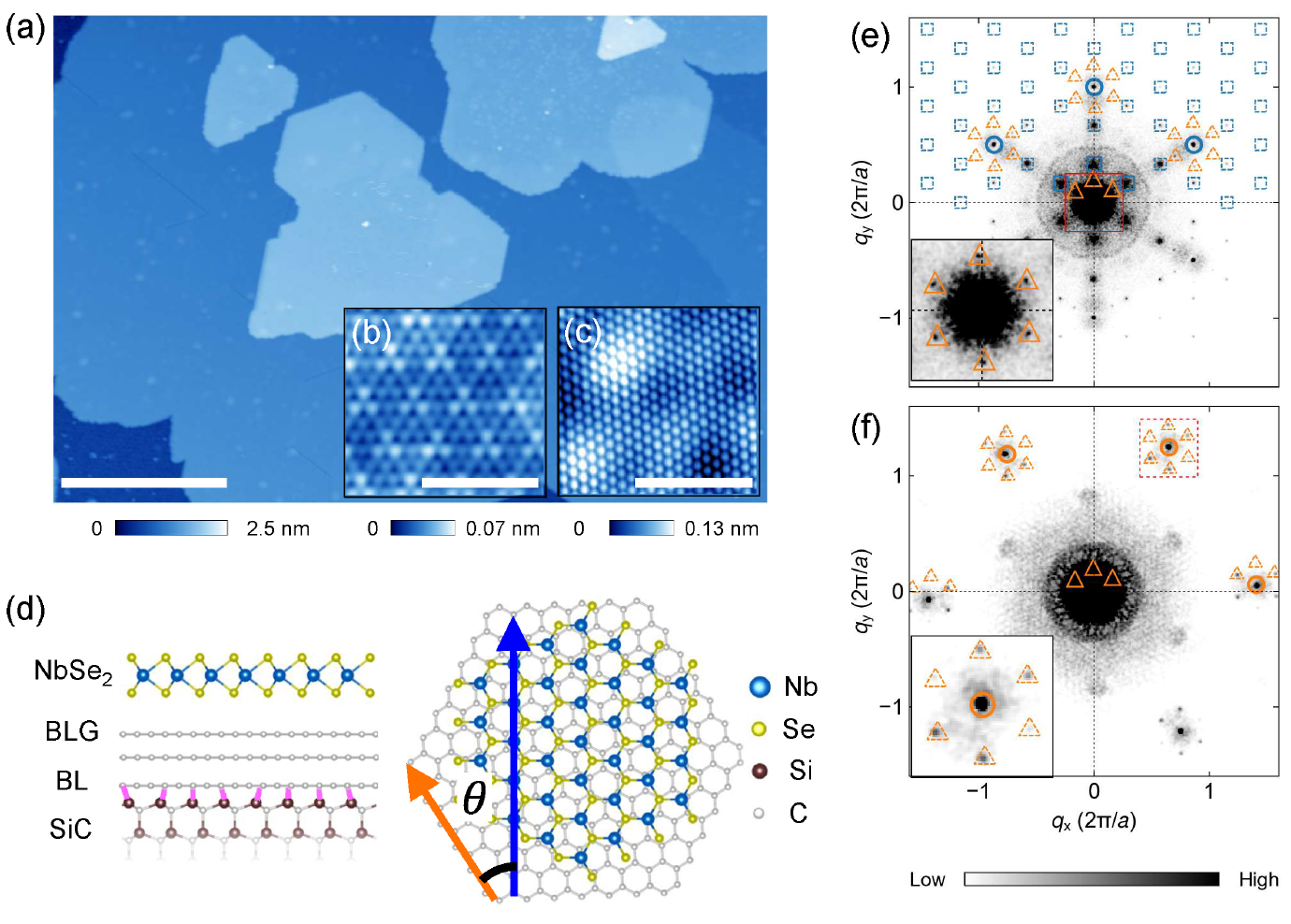}
\caption{
\textbf{STM characterizations of \ce{NbSe2} atomic layers grown on graphene.} 
(\textbf{a})
A typical large-scale topographic image of MBE-grown \ce{NbSe2} layers on graphene, with a scale bar indicating \SI{50}{\nm}.
The image was taken with a feedback set-point current $I_\mathrm{set} = \SI{100}{\pA}$ at a sample bias voltage $V_\mathrm{set} = \SI[retain-explicit-plus]{+0.5}{\V}$.
(\textbf{b})
A magnified topographic image of the monolayer \ce{NbSe2} showing atomic corrugations and a $3 \times 3$ CDW.
The scale bar denotes \SI{2.5}{\nm}.
Scanning conditions were $I_\mathrm{set} = \SI{100}{\pA}$ and $V_\mathrm{set} = \SI[retain-explicit-plus]{+3}{\mV}$.
(\textbf{c})
A magnified topographic image of the graphene region showing atomic corrugations and the $6\sqrt{3}\times6\sqrt{3}-\mathrm{R}\ang{30}$ DB order. The scale bar denotes \SI{2.5}{\nm}.
Scanning conditions were $I_\mathrm{set} = \SI{530}{\pA}$ and $V_\mathrm{set} = \SI[retain-explicit-plus]{+200}{\mV}$.
(\textbf{d})
Schematics of the side and top views of the \ce{NbSe2} monolayer on graphene, indicating magenta lines for bonds between the buffer layer graphene and the SiC substrate, forming the $6\sqrt{3}\times6\sqrt{3}-\mathrm{R}\ang{30}$ DB order.
The twist angle is defined as shown in the top view.
(\textbf{e} and \textbf{f})
Fourier-transformed topographic images of the \ce{NbSe2} monolayer and the graphene substrate, respectively.
We define $q_x$ and $q_y$ as $\mathbf{q}$ components in the Brillouin zone of \ce{NbSe2} parallel to the $\Gamma$-K and $\Gamma$-M directions, respectively.
$a = \SI{0.34}{nm}$ is the lattice constant of \ce{NbSe2}.
Symbols denoted by solid lines represent the fundamental $\mathbf{q}$ vectors of the modulations and those denoted by dashed lines signify higher harmonics and/or combinations of two different modulations.
Blue circles and squares denote the Bragg peaks of the Se atoms and the 3$\times$3 CDW peaks associated with the \ce{NbSe2}.
Orange symbols represent graphene-related peaks, including circles for graphene Bragg peaks and triangles for DB-order peaks.
Insets magnify regions highlighted by the dashed red squares.
}
\label{fig:1}
\end{figure*}

\clearpage
\newgeometry{top=0.0cm, left=1.0cm, right=1.0cm}
\begin{figure*}
\centering
\includegraphics[width=0.75\linewidth]{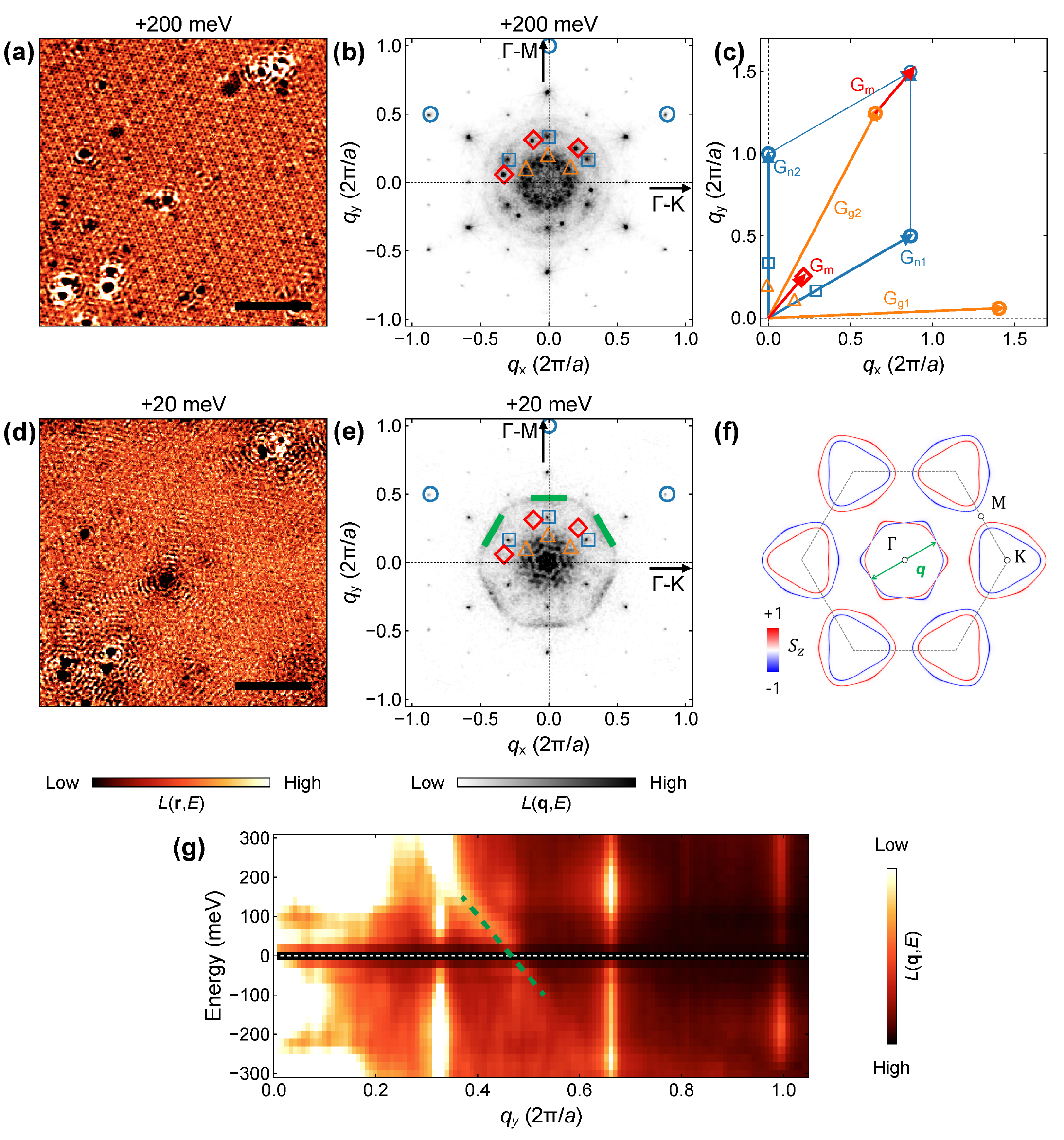}
\caption{\textbf{Moir\'e and QPI modulations in the \ce{NbSe2} monolayer on graphene with $\theta = \ang{28}$.}
(\textbf{a}) 
The normalized conductance map $L(\mathbf{r},\SI[retain-explicit-plus]{+200}{\meV})$.
The measurement conditions were $I_\mathrm{set} = \SI{200}{\pA}$, $V_\mathrm{set} = \SI[retain-explicit-plus]{+300}{\mV}$, and the bias modulation amplitude $V_\mathrm{mod} = \SI{20}{\mV}$.
The scale bar denotes \SI{10}{\nm}.
(\textbf{b}) 
$L(\mathbf{q},\SI[retain-explicit-plus]{+200}{\meV})$ map obtained by Fourier transforming (a).
Red diamonds denote the moir\'e vectors.
Other symbols have the same meaning as those in Fig.~\ref{fig:1}~(e and f).
(\textbf{c})
Relation among Bragg vectors of the atomic lattice of \ce{NbSe2} ($\mathbf{G}_\mathrm{n1}$ and $\mathbf{G}_\mathrm{n2}$ shown in blue), Bragg vectors of the graphene lattice ($\mathbf{G}_\mathrm{g1}$ and $\mathbf{G}_\mathrm{g2}$ shown in orange), and the moir\'e vector ($\mathbf{G}_\mathrm{m}$ shown in red) in $\mathbf{q}$ space.
In this particular case, $\mathbf{G}_\mathrm{m} = \mathbf{G}_\mathrm{n1}+\mathbf{G}_\mathrm{n2}-\mathbf{G}_\mathrm{g2}$.
In total, symmetrically equivalent six moir\'e peaks appear in $\mathbf{q}$ space.
(\textbf{d}) 
The normalized conductance image $L(\mathbf{r},\SI[retain-explicit-plus]{+20}{\meV})$ showing QPI modulations near defects.
The measurement conditions were the same as those for (a).
(\textbf{e}) 
$L(\mathbf{q},\SI[retain-explicit-plus]{+20}{\meV})$ obtained by Fourier transforming (d).
Green segments denote the QPI signals due to the intra-band scattering within the $\Gamma$-centered pocket shown in (f).
(\textbf{f})
Spin-resolved spectral function of the \ce{NbSe2} monolayer at $E = \SI[retain-explicit-plus]{+20}{\meV}$ calculated based on the tight-binding model derived from the density-functional-theory band structure~\cite{Liu2013}.
The dashed hexagon shows the first Brillouin zone.
$S_z$ denotes the $z$ component of the spin expectation value.
The green arrow denotes the scattering vector observed in (e).
(\textbf{g})
Line profiles of $L(\mathbf{q},E)$ along the $\Gamma$-M direction in $\mathbf{q}$ space.
The hole-like QPI dispersion is shown by the green dashed line.
Three dispersionless features correspond to the CDW vector, its harmonics, and the lattice Bragg vector of \ce{NbSe2}.
}
\label{fig:2}
\end{figure*}

\clearpage
\newgeometry{left=1.0cm, right=1.0cm}
\begin{figure*}
\centering
\includegraphics[width=0.9\linewidth]{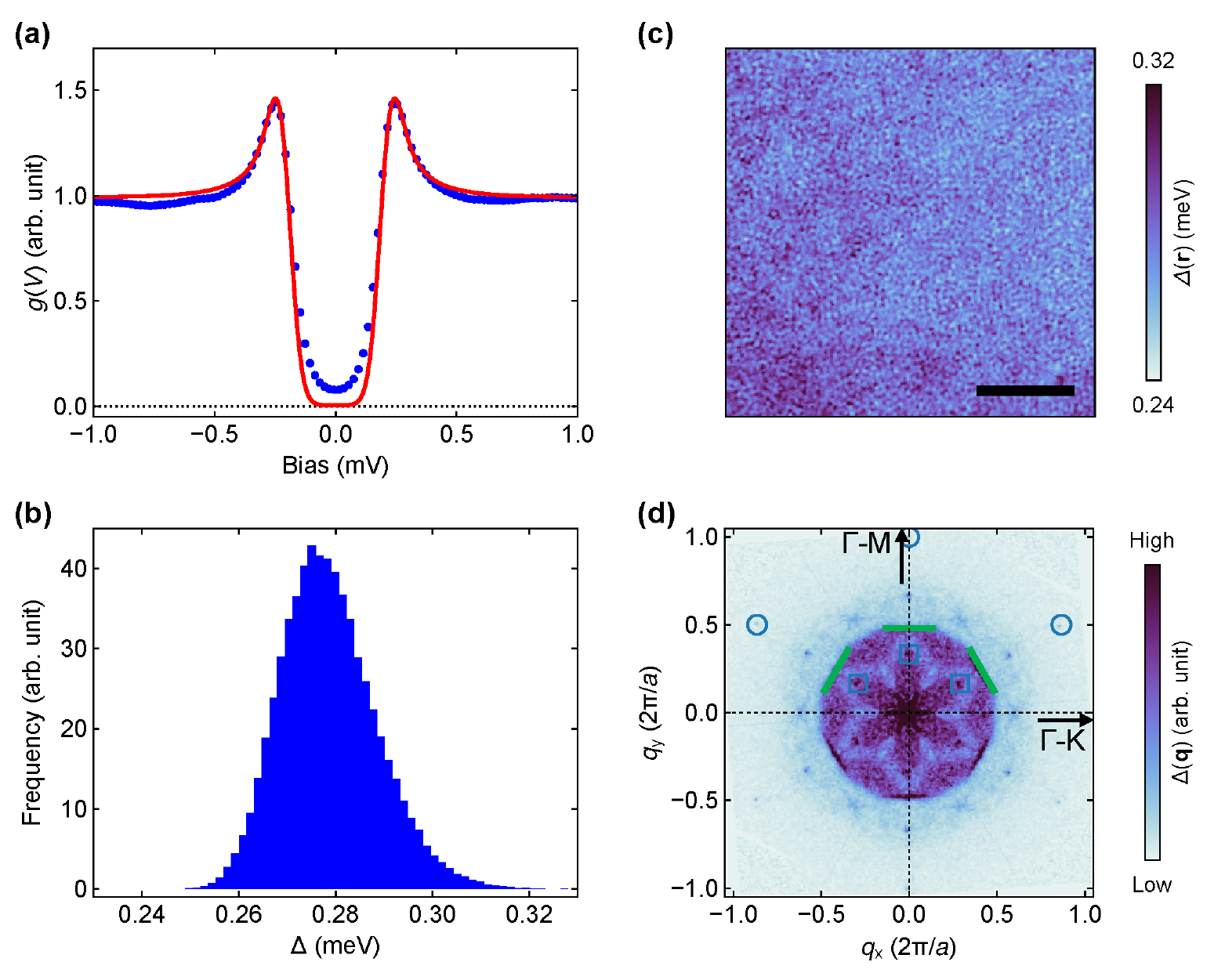}
\caption{
\textbf{Superconducting gap of the \ce{NbSe2} monolayer on graphene with $\theta = \ang{28}$.}
(\textbf{a}) 
Tunneling spectrum averaged over the $\SI{50}{\nm} \times \SI{50}{\nm}$ field of view, showing the superconducting gap.
The measurement conditions were $I_\mathrm{set} = \SI{200}{\pA}$, $V_\mathrm{set} = \SI[retain-explicit-plus]{+3}{\mV}$, and the bias modulation amplitude $V_\mathrm{mod} = \SI{15}{\micro\eV}$.
The red line denotes the spectrum expected from the DCB model (see the main text and the Supplementary Materials).
(\textbf{b}) 
Histogram of the gap amplitude $\Delta(\mathbf{r})$ measured over the $\SI{50}{\nm} \times \SI{50}{\nm}$ field of view.
(\textbf{c}) 
Spatial distribution of the superconducting gap amplitude $\Delta(\mathbf{r})$. The scale bar denotes 10 nm.
(\textbf{d}) 
Fourier transformed image of $\Delta(\mathbf{r})$ showing the intra-band scattering with the $\Gamma$-centered pocket signals (green segments) and the 3$\times$3 CDW peaks (blue squares).
}
\label{fig:3}
\end{figure*}

\clearpage
\newgeometry{top=1.0cm, left=1.0cm, right=1.0cm}
\begin{figure*}
\centering
\includegraphics[width=\linewidth]{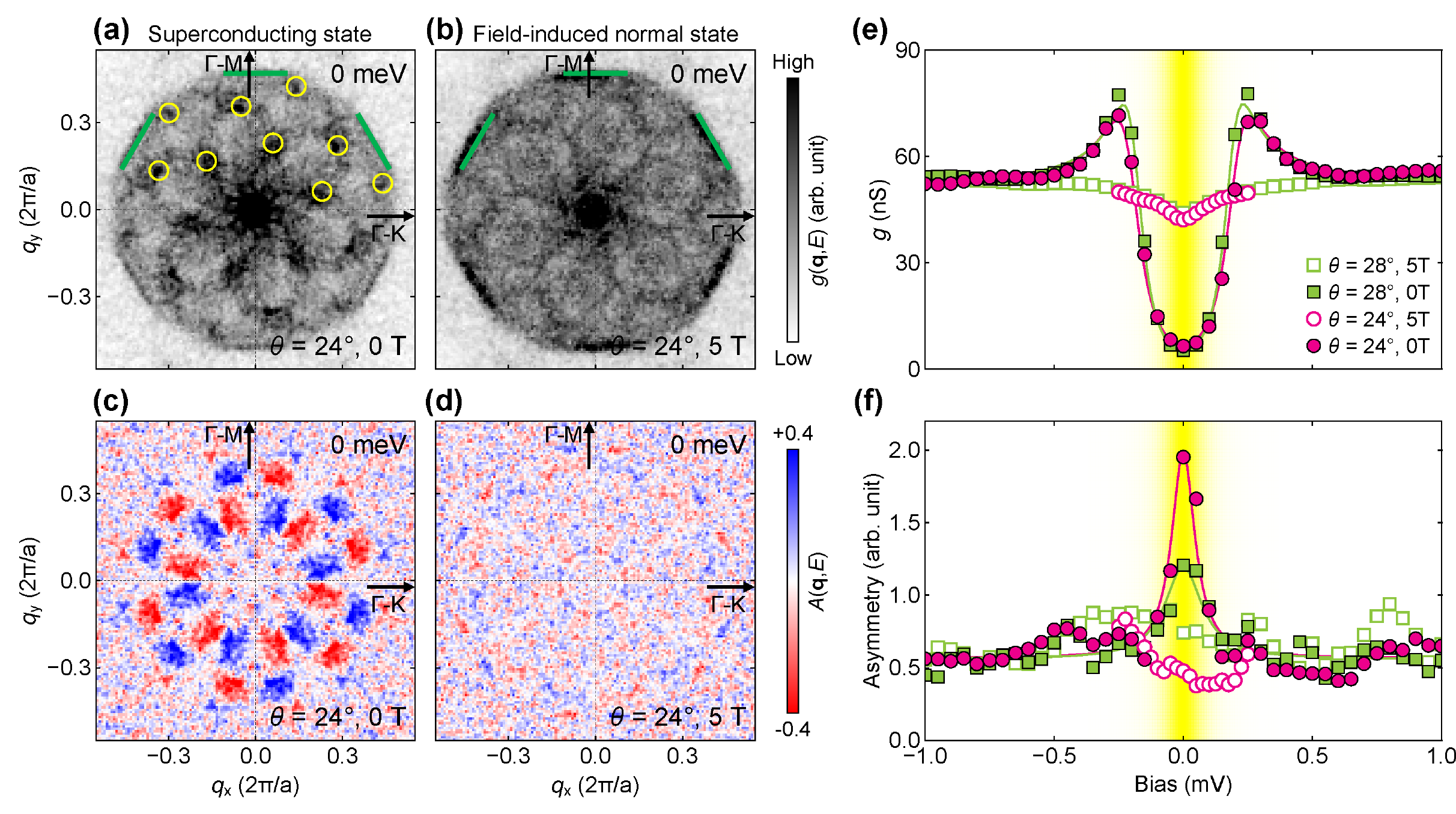}
\caption{
\textbf{Twisted Bogoliubov quasiparticles.} 
(\textbf{a})
$g(\mathbf{q},\SI{0}{meV})$ map in the superconducting state for the sample with $\theta = \ang{24}$, showing the $\mathbf{q}$ components of the residual LDOS modulations at the Fermi energy.
Green segments denote the QPI signals due to the intra-band scattering within the $\Gamma$-centered pocket.
Yellow circles show signals that break the mirror symmetry with respect to the $\Gamma$-M and $\Gamma$-K lines of the \ce{NbSe2} monolayer.
The original $g(\mathbf{q},\SI{0}{meV})$ map before the Fourier transformation was measured in the $\SI{38}{\nm} \times \SI{38}{\nm}$ field of view with conditions $I_\mathrm{set} = \SI{200}{\pA}$, $V_\mathrm{set} = \SI[retain-explicit-plus]{+3}{\mV}$, and $V_\mathrm{mod} = \SI{50}{\micro\eV}$.
(\textbf{b})
$g(\mathbf{q},\SI{0}{meV})$ map in the normal state under a magnetic field $\mu_0H = \SI{5}{\tesla}$ perpendicular to the surface.
All other conditions are the same as those in (a).
All the characteristic $\mathbf{q}$ vectors except for the QPI signals (green segments) disappear, signifying their superconducting origin.
(\textbf{c})
Chiral component map $A(\mathbf{q},\SI{0}{meV})$ in the superconducting state showing the $\mathbf{q}$ components that do not respect the mirror symmetry of \ce{NbSe2}.
(\textbf{d})
Same as (c) but in the normal state where the chiral nature disappears.
(\textbf{e} and \textbf{f})
Tunneling spectra averaged over the fields of view and the energy dependence of the degree of chirality obtained by integrating the absolute value of $A(\mathbf{q},E)$ over the $\mathbf{q}$ range studied, respectively.
Results for the samples with two different twisting angles are shown.
The chiral nature is seen only inside the superconducting gap and disappears in the normal state.
}
\label{fig:4}
\end{figure*}

\clearpage
\newgeometry{top=1.0cm, left=1.0cm, right=1.0cm}
\begin{figure*}
\centering
\includegraphics[width=\linewidth]{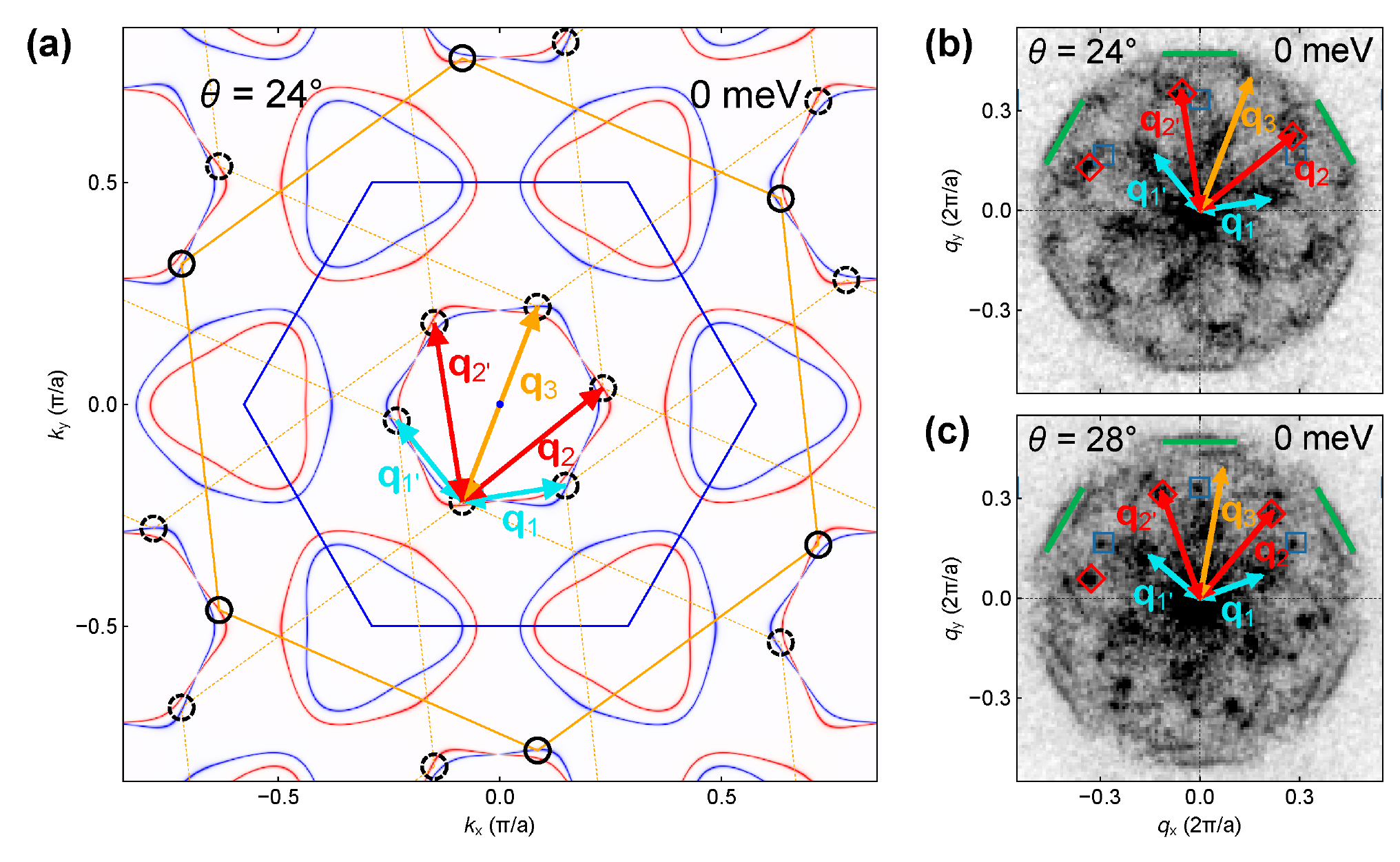}
\caption{
\textbf{Sextet model of the twisted Bogoliubov QPI.} 
(\textbf{a})
The momentum-space electronic state of the twisted stack of \ce{NbSe2} monolayer on graphene with $\theta = \ang{24}$.
Blue and orange hexagons represent the Brillouin zones of the \ce{NbSe2} monolayer and graphene, respectively.
Spin-resolved spectral function of \ce{NbSe2} is shown in the same manner as Fig.~\ref{fig:2}(f) but in the extended zone scheme.
The solid black circles represent the Fermi surface of the innermost graphene's valleys. The graphene Fermi surfaces shifted by the Bragg vectors of \ce{NbSe2} are shown by the dashed black circles.
We assume that the Fermi energy of graphene is 0.4 eV lower than the Fermi energy of \ce{NbSe2}~\cite{Coletti2010,Gani2019}.
The arrows indicate the possible Bogoliubov QPI vectors connecting the sextet regions where the Fermi surfaces of \ce{NbSe2} and graphene overlap.
(\textbf{b})
Observed Bogoliubov QPI pattern for $\theta = \ang{24}$.
All the vectors expected in (a) are identified.
Green segments denote the QPI signals due to the intra-band scattering within the $\Gamma$-centered pocket shown in Fig.~\ref{fig:2}(f).
Red diamonds and blue squares show the moir\'e vectors and the 3$\times$3 CDW peaks, respectively.
(\textbf{c})
Same as (b) but for $\theta = \ang{28}$.
}
\label{fig:5}
\end{figure*}
\clearpage

\appendix
\restoregeometry
\setcounter{figure}{0}
\renewcommand{\thefigure}{S\arabic{figure}}
\newcounter{SIfig}
\renewcommand{\theSIfig}{S\arabic{SIfig}}
\section{Materials and Methods}\label{materials-and-methods}

\subsection{MBE growth of the \texorpdfstring{\ce{NbSe2}}{NbSe2} monolayer on 4\emph{H}-SiC(0001)}

The \ce{NbSe2} monolayer films were grown on
bilayer-graphene-terminated 4\emph{H}-SiC(0001) substrates using MBE. We
utilized a commercially available $n$-type SiC substrate with a typical
resistivity of 0.01 \(\Omega\)cm. The epitaxial graphene surfaces were
obtained by annealing the substrate at approximately 1500${}^\circ$C in the
MBE chamber with a base pressure of 3×10\textsuperscript{-7}~Pa. After
the substrate was cooled to the growth temperature of 490${}^\circ$C,
high-purity Nb and Se were evaporated using a rod-type electron beam
evaporator and a standard Knudsen cell, respectively. The flux ratio of
Nb to Se was approximately 1:15. After the growth, the samples were
post-annealed in situ at 450${}^\circ$C for 20 minutes to enhance the
crystalline quality.

\subsection{SI-STM measurements and analysis}

The SI-STM measurements were performed using a homemade
dilution-refrigerator-based ultra-high-vacuum (UHV) STM system with the
base electron temperature of about 90 mK~\cite{Machida2018}. All SI-STM
measurements were conducted at the base temperature. The samples
prepared in the separate MBE chamber were transferred to the STM chamber
using a homemade UHV suitcase evacuated by the battery-operated ion
pump. The pressure during the transfer was kept at about
1×10\textsuperscript{-8}~Pa. We utilized an electrochemically etched
tungsten wire as a scanning tip, which was cleaned by field evaporation
followed by conditioning on a clean Cu(111) surface. All the STM
topographic images were taken with constant current scanning.

The differential conductance spectrum was measured using a standard
lock-in technique with a modulation frequency of 614.7 Hz. We carefully
chose the fields of view for SI-STM to be a single domain clean
\ce{NbSe2} monolayer. To mitigate the effect of the drift
during the scanning and enhance the signal-to-noise ratio, all the FT
images obtained by spectroscopic scan were affine transformed in
\textbf{q} space and symmetrized according to the
\emph{C}\textsubscript{3} symmetry, which is the lowest symmetry
expected in the twisted stack of \ce{NbSe2} monolayer and
graphene.

\section{Supplementary Text}
\subsection{Fitting tunneling spectra with the DCB model}

\begin{figure*}[ht]
\centering
\includegraphics[width=\linewidth]{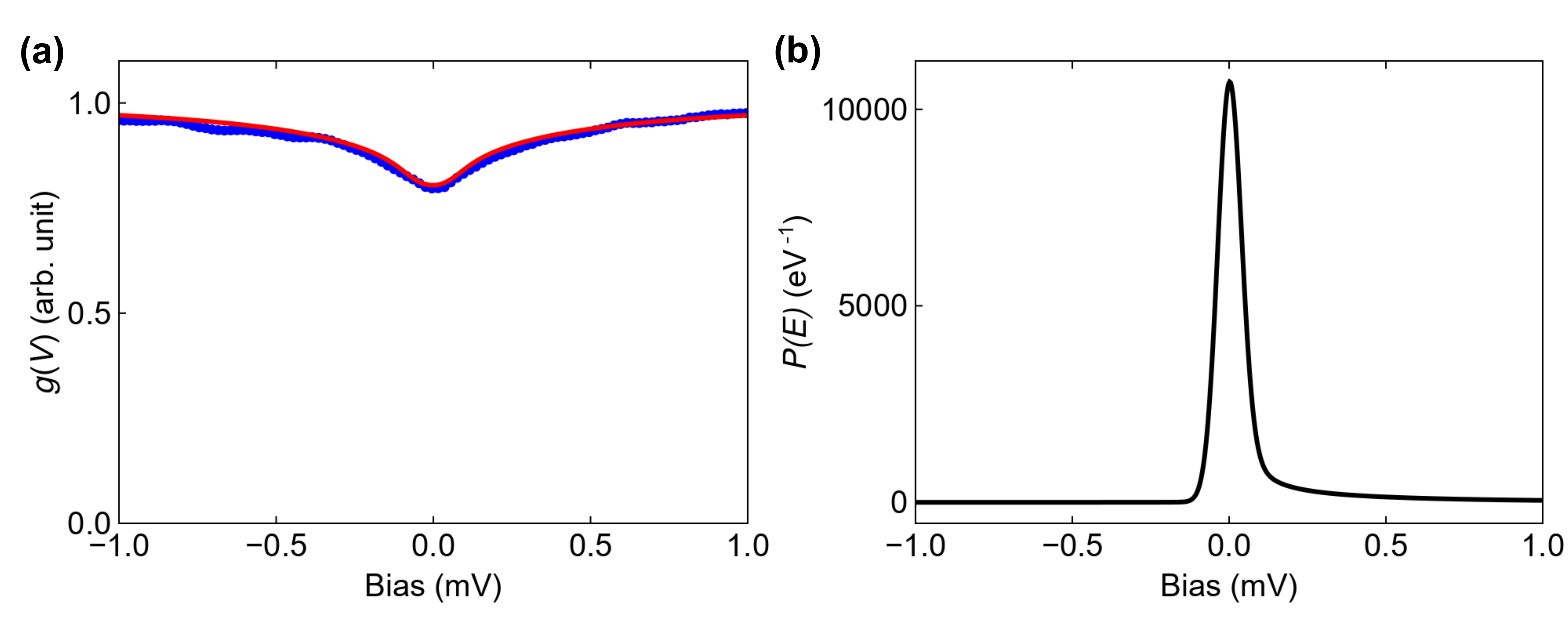}
\caption{
(\textbf{a})
  Tunneling spectrum averaged over the same field of view as Fig.~\ref{fig:3}(a) of
  the main text in the normal state under a magnetic field
  \(\mu_{0}H =\) 5 T perpendicular to the surface. The measurement
  conditions were the same as those in Fig.~\ref{fig:3}(a) of the main text. The red
  line denotes the spectrum obtained by the fitting based on the DCB
  model.
(\textbf{b})
  The \(P(E)\) function obtained by fitting the tunneling spectrum in
  the field-induced normal state.
}
\refstepcounter{SIfig}\label{fig:S1}
\end{figure*}

The energy exchange between tunneling electrons and the environment
results in the DCB effect, which becomes prominent in the tunneling
experiment on island-like thin films deposited on poorly conducting
substrates~\cite{Brun2012}, as is the case of the present experiment. The
DCB effect can be described by the so-called \(P(E)\) function~\cite{Devoret1990, Ingold1992}. The tunneling spectrum in the presence of the DCB
effect is the LDOS spectrum convoluted by \(P(E)\), showing a
characteristic cusp-like dip near \emph{V} = 0 in the normal state with
the energy-independent LDOS~\cite{Brun2012,Ast2016}. The \(P(E)\) function can
be evaluated by analyzing this dip. However, in our superconducting
film, the superconducting gap masks the dip. Therefore, we suppressed
superconductivity by applying a high magnetic field of 5 T perpendicular
to the surface. As shown in Fig.~\ref{fig:S1}(a), the observed spectrum in the
field-induced normal state exhibits a cusp-like dip at \emph{V} = 0,
suggesting that the DCB effect matters.

Following Ref.\cite{Ast2016}, we express the \(P(E)\) function using the
principal resonance frequency \(\omega_{0}\), the junction capacitance
\(C_{J}\), the damping parameter \(\alpha\), and the junction
temperature \(T\) and evaluate these parameters by fitting the observed
tunneling spectrum in the field-induced normal state. To simplify the
fitting process, we set the junction temperature to be the previously
estimated electron temperature of \(T\) = 90 mK~\cite{Machida2018}. The fitting
was performed by searching for the optimal parameter set of
\(\omega_{0}\), \(C_{J}\), and \(\alpha\) in the three-dimensional space
of \(0.1 < {\hbar\omega}_{0} < 100\ \mu eV\),
\(0.1 < C_{J} < 100\ fF\), and \(0 < \alpha < 1\). The obtained
parameters were \(\hbar\omega_{0}\) = 7.0 \(\mu\)eV, \(C_{J}\) = 1.0
fF, and \(\alpha\) = 0.75. The \(P(E)\) function calculated with these
parameters is shown in Fig.~\ref{fig:S1}(b). We also tried the same procedure with
\(T\) = 40 mK, which is the thermometer temperature, and an equally good
fit was achieved with slightly different parameters for \(\omega_{0}\),
\(C_{J}\), and \(\alpha\).

Using the \(P(E)\) function obtained above, we convolve the model
superconducting gap function and compare the result with the observed
spectrum shown in Fig.~\ref{fig:3}(a) of the main text. Here, we assume that
\(P(E)\) does not depend on the magnetic field and that the LDOS
spectrum of the field-induced normal state is energy independent. We
adopt the simplest Bardeen-Cooper-Schrieffer (BCS) model for the LDOS in
the superconducting state. Here, the gap magnitude is the only
adjustable parameter. As shown by the red line in Fig.~\ref{fig:3}(a) of the main
text, this simple model well reproduces the broadened gap-edge peaks.
However, the observed significant residual spectral weight near \emph{V}
= 0 indicates that the actual LDOS spectrum has more low-energy states
than the BCS gap function.

\subsection{Superconducting gap map analysis}

We analyze the spatial variation of the gap amplitude,
\(\Delta(\mathbf{r})\), defined as half the energy separation between
the peaks at the gap edge. To evaluate the peak energy, we divide the
tunneling spectra into positive and negative energy segments and fit
each of them phenomenologically with a skewed Lorentzian,

\[f_{SL}(x;x_{0},\gamma,a)\  = \ \frac{1}{\left\lbrack \pi\gamma\left\lbrack 1 + \frac{\left| x - x_{0} \right|^{2}}{\gamma^{2}\left( 1 + a\ sgn\left( x - x_{0} \right) \right)^{2}} \right\rbrack \right\rbrack},\ \ \]

where \(x_{0}\) is the location parameter, \(\gamma > 0\) is the scale
parameter, and $- 1 < a < 1$ is the skewness parameter. The obtained
\(\Delta(\mathbf{r})\) map is shown in Fig.~\ref{fig:3}(c).

\subsection{Magnetic field dependence of the tunneling spectrum}

\begin{figure*}[ht]
\centering
\includegraphics[width=\linewidth]{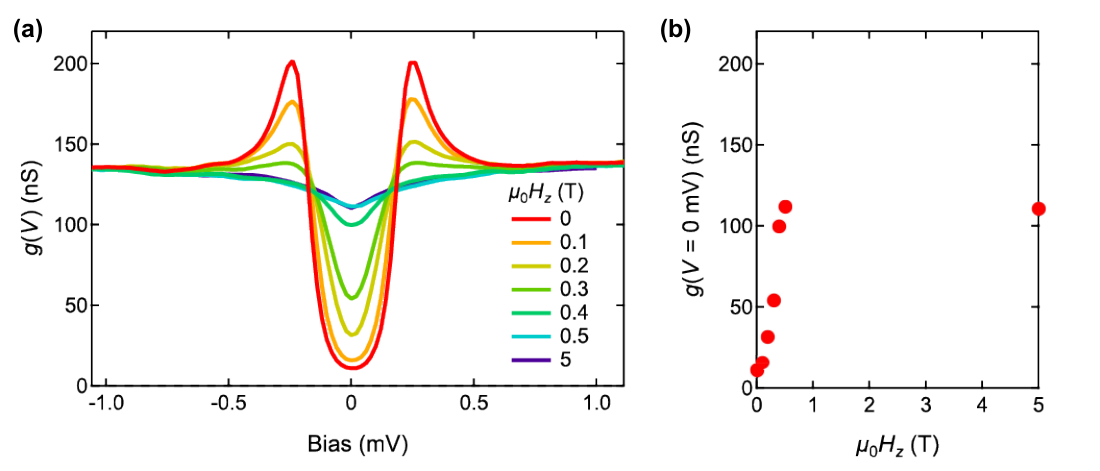}
\caption{
(\textbf{a})
  Magnetic-field dependence of the tunneling spectra \(g(V)\) for
  \(\theta\) = 28\({^\circ}\). The measurement conditions were
  \(I_{set}\) = 200 pA, \(V_{set}\) = +3 mV, and the bias modulation
  amplitude \(V_{mod}\) = 15~\(\mu\)eV.
(\textbf{b})
  Magnetic-field dependence of \(g(V)\) at the Fermi energy.
}
\refstepcounter{SIfig}\label{fig:S2}
\end{figure*}

To evaluate the upper critical field of the \ce{NbSe2}
monolayer on graphene, we examine the effect of magnetic field on the
tunneling spectrum. Figure ~\ref{fig:S2}(a) depicts the spectra of the monolayer with
\(\theta\) = 28\({^\circ}\) at different magnetic fields. Each spectrum
is spatially averaged over a 40~nm × 40~nm area. We also show the field
dependence of \(g(V)\) at the Fermi energy in Fig.~\ref{fig:S2}(b). The
superconducting gap is gradually suppressed by the magnetic field, and
the spectra above 0.5 T are essentially field-independent. Therefore,
the upper critical field on this monolayer is about 0.5~T and the
cusp-like dip at the Fermi energy at a higher field is likely to be due
to the DCB effect.

\subsection{Twist angle dependence of the superconducting gap}
\begin{figure*}[ht]
\centering
\includegraphics[width=0.6\linewidth]{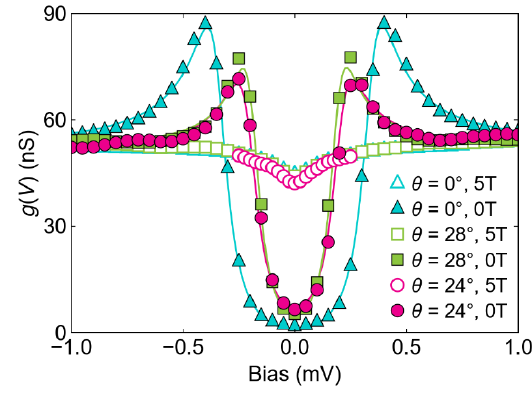}
\caption{
The superconducting-gap spectra of the \ce{NbSe2} monolayers
on graphene with different twist angles.
}
\refstepcounter{SIfig}\label{fig:S3}
\end{figure*}

\begin{figure*}[ht]
\centering
\includegraphics[width=0.9\linewidth]{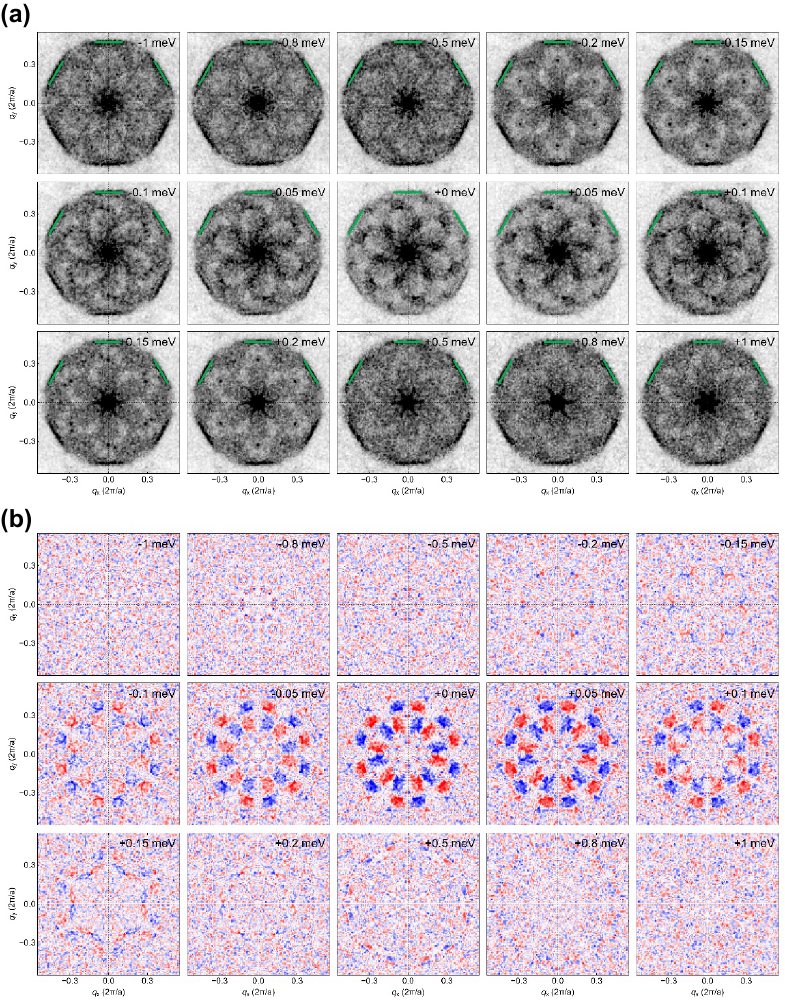}
\caption{
(\textbf{a})
  \(g(\mathbf{q},E)\) map in the superconducting state for the sample
  with \(\theta\) = 24\({^\circ}\) at different energies. Green segments
  denote the BQPI signals due to the intra-band scattering within the
  \(\Gamma\)-centered pocket.
(\textbf{b})
  Chiral component map \(A(\mathbf{q},E)\) in the superconducting state
  for the sample with \(\theta\) = 24\({^\circ}\) at different energies.
}
\refstepcounter{SIfig}\label{fig:S4}
\end{figure*}

\begin{figure*}[ht]
\centering
\includegraphics[width=0.9\linewidth]{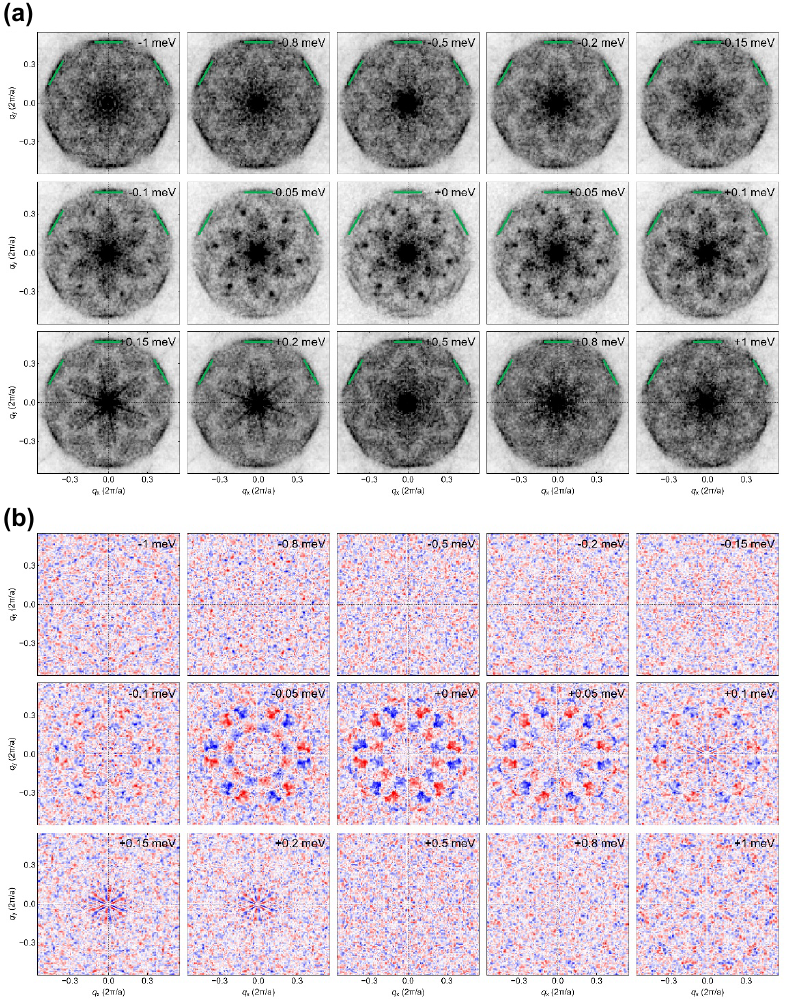}
\caption{
(\textbf{a})
  \(g(\mathbf{q},E)\) map in the superconducting state for the sample
  with \(\theta\) = 28\({^\circ}\) at different energies. Green segments
  denote the BQPI signals due to the intra-band scattering within the
  \(\Gamma\)-centered pocket.
(\textbf{b})
  Chiral component map \(A(\mathbf{q},E)\) in the superconducting state
  for the sample with \(\theta\) = 28\({^\circ}\) at different energies.
}
\refstepcounter{SIfig}\label{fig:S5}
\end{figure*}

\begin{figure*}[ht]
\centering
\includegraphics[width=0.9\linewidth]{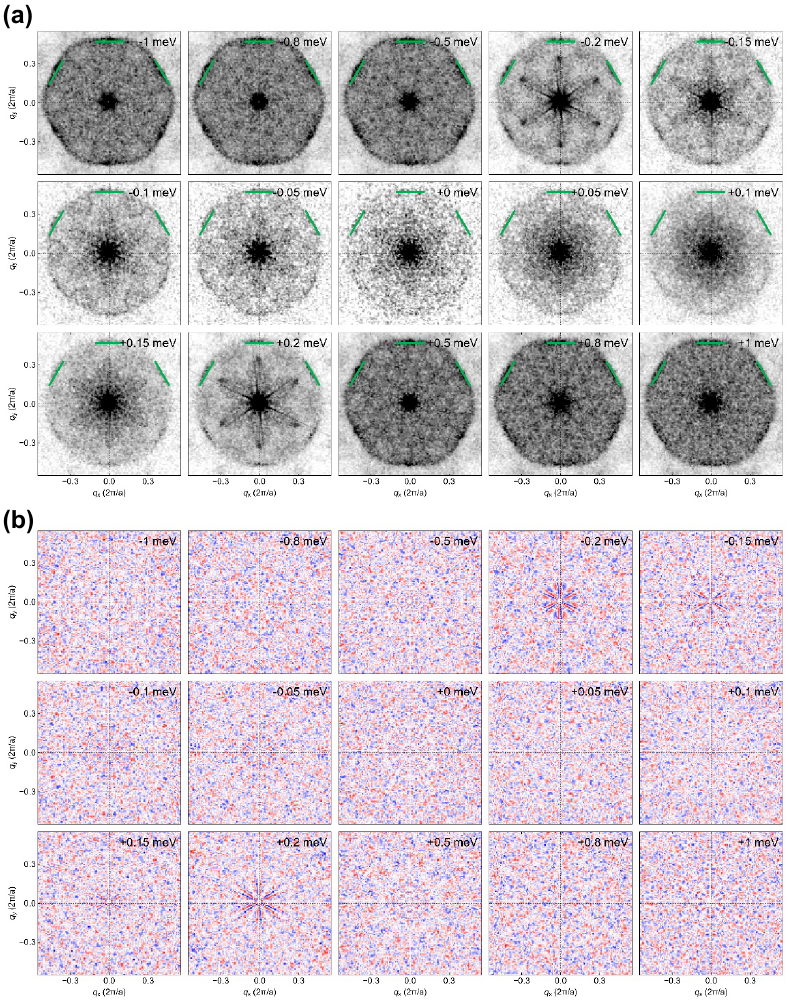}
\caption{
(\textbf{a})
  \(g(\mathbf{q},E)\) map in the superconducting state for the sample
  with \(\theta\) = 0\({^\circ}\) at different energies. Green segments
  denote the BQPI signals due to the intra-band scattering within the
  \(\Gamma\)-centered pocket.
(\textbf{b})
  Chiral component map \(A(\mathbf{q},E)\) in the superconducting state
  for the sample with \(\theta\) = 0\({^\circ}\) at different energies.
}
\refstepcounter{SIfig}\label{fig:S6}
\end{figure*}

The spatially averaged tunneling spectra for the sample with   \(\theta\) = 24, 28, and 0\({^\circ}\) are shown in Fig.~\ref{fig:S3}. The selected energy slices for the \(g(\mathbf{q},E)\) map and corresponding chiral component map  \(A(\mathbf{q},E)\) are shown in Fig.~\ref{fig:S4} to Fig.~\ref{fig:S6}.

\clearpage

\clearpage

\clearpage

\end{document}